\documentclass[12pt,reqno]{amsart}
\renewcommand{\baselinestretch}{1.26}
\usepackage{epsfig}
\usepackage{rotating}
\usepackage{amssymb}
\usepackage{amsthm}
\usepackage{multirow}
\usepackage{lscape}
\usepackage{graphicx}
\usepackage{verbatim} 
\makeatletter
\newcommand{\singlespacing}{\let\CS=\@currsize\renewcommand{\baselinestretch}{1}\tiny\CS}
\begin{document}

\parskip = 10pt
\def \qed {\hfill \vrule height7pt width 5pt depth 0pt}
\newcommand{\ve}[1]{\mbox{\boldmath$#1$}}
\newcommand{\IR}{\mbox{$I\!\!R$}}
\newcommand{\1}{\Rightarrow}
\newcommand{\bs}{\baselineskip}
\newcommand{\esp}{\end{sloppypar}}
\newcommand{\be}{\begin{equation}}
\newcommand{\ee}{\end{equation}}
\newcommand{\beanno}{\begin{eqnarray*}}
\newcommand{\inp}[2]{\left( {#1} ,\,{#2} \right)}
\newcommand{\eeanno}{\end{eqnarray*}}
\newcommand{\bea}{\begin{eqnarray}}
\newcommand{\eea}{\end{eqnarray}}
\newcommand{\ba}{\begin{array}}
\newcommand{\ea}{\end{array}}
\newcommand{\nno}{\nonumber}
\newcommand{\dou}{\partial}
\newcommand{\bc}{\begin{center}}
\newcommand{\ec}{\end{center}}
\newcommand{\2}{\subseteq}
\newcommand{\cl}{\centerline}
\newcommand{\boxeq}[2]

\def\refhg{\hangindent=20pt\hangafter=1}
\def\refmark{\par\vskip 2.50mm\noindent\refhg}

\def \qed {\hfill \vrule height7pt width 5pt depth 0pt}

\def\lam{\lambda }
\def\Lam{\Lambda}
\def\lab{\label }
\def\iomui{[0,1/ \mu)}
\def\komu{\frac{k}{\mu}}
\def\kmu{\frac{k}{\mu}}
\def\Komu{\frac{K}{\mu}}
\def\alst{\alpha\sim t}
\def\tal{{\tau}^{\al}}
\def\abal{\mid{\alpha}\mid}
\def\la{\langle}
\def\ra{\rangle}
\def\rar{\rightarrow}
\def\Rar{\Rightarrow}
\def\ho{\hat{\omega}}
\def\to{\tilde{\omega}}
\def\o{\omega}

\def\cA{{\cal A}}
\def\cB{{\cal B}}
\def\cC{{\cal C}}
\def\cD{{\cal D}}
\def\cE{{\cal E}}
\def\cF{{\cal F}}
\def\cG{{\cal G}}
\def\cH{{\cal H}}
\def\cI{{\cal I}}
\def\cJ{{\cal J}}
\def\cK{{\cal K}}
\def\cL{{\cal L}}
\def\cM{{\cal M}}
\def\cN{{\cal N}}
\def\cO{{\cal O}}
\def\cP{{\cal P}}
\def\cQ{{\cal Q}}
\def\cR{{\mathcal R}}
\def\cS{{\cal S}}
\def\cT{{\cal T}}
\def\cU{{\cal U}}
\def\cV{{\cal V}}
\def\cW{{\cal W}}
\def\cX{{\cal X}}
\def\cY{{\cal Y}}
\def\cZ{{\cal Z}}

\def\1N{\frac{1}{n}}
\def\lb{\lbrace}
\def\rb{\rbrace}
\def\blp{\bigl (}
\def\brp{\bigr )}
\def\blb{\bigl \lbrace}
\def\brb{\bigr \rbrace}
\def\bls{\bigl [ }
\def\brs{\bigr ] }
\def\Blp{\Bigl (}
\def\Brp{\Bigr )}
\def\Blb{\Bigl \lbrace}
\def\Brb{\Bigr \rbrace}
\def\BLB{\Biggl \lbrace}
\def\BRB{\Biggr \rbrace}
\def\Bls{\Bigl [ }
\def\Brs{\Bigl ] }
\def\BLS{\Biggl [ }
\def\BRS{\Biggl ] }
\def\Skor{D([0, 1], \cS' (R^d))}
\def\mlnm{(m\lambda+n\mu)}
\def\ds{\displaystyle}
\def\what{\widehat}
\def\wtilde{\widetilde}

\def\limn{\lim_{n \rightarrow \infty}}
\def\limiN{\liminf_{n\rightarrow \infty}}
\def\stN{\sum_{t=1}^n}
\def\stn{\sum_{t=1}^n}
\def\sjp{\sum_{j=1}^p}
\def\skp{\sum_{k=1}^p}
\def\skii{\sum_{k=0}^\infty}
\def\slii{\sum_{l=0}^\infty}

\def\ptN{\prod_{t=1}^n}
\def\ptn{\prod_{t=1}^n}
\def\wh{\widehat}
\def\wt{\widetilde}
\def\a012{a_0 + a_1 t + a_2 t^2}
\def\0a012{a_0^0 + a_1^0 t + a_2^0 t^2}
\def\0t012{\theta_0^0 + \theta_1^0 t + \theta_2^0 t^2}
\def\t012{\theta_0 + \theta_1 t + \theta_2 t^2}
\def\blfootnote{\xdef\@thefnmark{}\@footnotetext} 
\theoremstyle{plain}
\newtheorem{theorem}{\bf \sc Theorem}[section]
\theoremstyle{definition}
\newtheorem{remark}{\bf Remark}
\theoremstyle{plain}
\newtheorem{lemma}{\bf Lemma}
\newtheorem{assum}{\bf Assumption}
\theoremstyle{definition}
\newtheorem{pth}{\bf Proof of Theorem}[]
\newtheorem{plm}{\bf Proof of Lemma}[]

\newcommand{\dotx}{\dot{X}}
\newcommand{\ddotx}{\ddot{X}}

\title[Chirp Signal with Heavy Tailed Error]{Estimation of Parameters of Multiple Chirp Signal in presence of Heavy Tailed Errors}
\author{} 
\author{Swagata Nandi$^1$}
\author{Debasis Kundu$^2$}
\address{$^1$Theoretical Statistics and Mathematics Unit, Indian Statistical Institute, 7, S.J.S. Sansanwal Marg, New Delhi - 110016,
India, nandi@isid.ac.in}
\address{$^2$Department of Mathematics and Statistics, Indian Institute of
Technology Kanpur, Pin 208016, India, kundu@iitk.ac.in} 


\begin{abstract}
In this paper, we consider the estimation of the unknown parameters of the multiple chirp signal model in presence of additive error.  
The chirp signals are quite common in many 
areas of science and engineering, specially sonar, radar, audio signals etc.  The observed signals are usually corrupted by noise.
In different signal processing applications it is observed that the errors may be heavy tailed.  In this paper it is assumed that the 
additive errors have mean zero but may not have 
finite variance and are independent and identically distributed.  We consider the least squares estimators and the approximate 
least squares estimators which maximize a periodogram like function. It has been observed that both the estimators are strongly 
consistent.  The asymptotic distribution of the least squares estimators is obtained under the assumption that the additive errors 
are from a symmetric stable distribution.  The approximate least squares estimators have the same asymptotic distribution as the 
least squares estimators.  We perform some numerical simulations to see how the proposed estimators work. It is observed that
the least squares estimators perform slightly better than the approximately least squares estimators in terms of the biases and
mean absolute deviation.
\end{abstract}

\maketitle

\section{Introduction}   \label{sec1}
This paper is concerned with the estimation of the unknown parameters of the following multicomponent chirp signal model;
\be
y(t) = \sum_{k=1}^p \Blp A_k^0 \cos(\theta_{1k}^0 t + \theta_{2k}^0 t^2) + B^0 \sin(\theta_{1k}^0 t + \theta_{2k}^0 t^2) \Brp
+ e(t); \hspace{.15in} 
t = 1, \ldots, n.   \quad \quad \label{multiple-chirp-model}
\ee
Here $y(t)$ is a real-valued signal observed at $n$ equi-distant points, $t=1,\ldots, n$.  For $k=1,\ldots, p$, $A_k^0$ and $B_k^0$ 
are real-valued amplitudes and $\theta_{1k}^0$ and $\theta_{2k}$ are frequency and frequency rate, respectively.  The number of chirp 
components, $p\ge1$, is assumed to be known in advance.   The additive error $\{e(t)\}$ is a sequence of independent and identically 
distributed (i.i.d.) random variables with mean zero but the variance may not exist.  This includes the case when the variance is 
not finite.  The specific structure of $\{e(t)\}$  will be stated in Assumptions \ref{assum1} and \ref{assum2}. 

Model \eqref{multiple-chirp-model} is a generalization of fixed-frequencies sinusoidal model.  In chirp signal model, frequencies 
change with time and the rate of change is governed by the frequency rate parameter in case of each chirp component.  This model is 
quite useful in various areas of science and engineering, specially, sonar, radar, communications etc.  For example, chirp 
signal model is used in various forms of trajectories of moving objects with respect to fixed receivers.  Several authors have 
considered this or some related models when $e(t)$'s are i.i.d. with finite variance, see for example Abatzoglou \cite{Abatz:1986}, 
Djuric and Kay \cite{DK:1990}, Giannakis and Zhou \cite{GZ:1995}, Shamsunder et al. \cite{SGF:1995}, 
Ikram et al. \cite{IAH:1997}, Gini et al. \cite{GMV:2000}, Saha and Kay \cite{SK:2002}, Nandi and Kundu \cite{NK:2004}, 
 Farquharson et al. \cite{FOL:2005}, 
Mazumder \cite{Mazumder:2017}, 
Jensen et al. \cite{JNJCJ:2017} and see the references cited therein.  

In several signal processing applications it is observed that the errors are heavy tailed, see for example Zoubir and Brcich 
\cite{ZB:2002} and Barkat and Stankovic \cite{BS:2004}.   The main aim of this paper is to consider the case when the error random 
variables have a heavy tailed distribution.  In a heavy tailed distribution, extreme probabilities approach zero relatively slowly.  
An important criterion for heavy tailedness, as noted by Mandelbrot \cite{MANDEL:1963}, is the non-existence of the variance.  We 
are using the same definition as of Mandelbrot \cite{MANDEL:1963} that a distribution is heavy tailed if and only if the variance is 
not finite.  It is assumed that $E|e(t)|^{1+\delta} < \infty$, for $0<\delta <1$.  It may be mentioned that the  
LS method is a reasonable choice for estimating the unknown parameters in any linear and nonlinear models in presence of additive 
errors.  But when the second moment of the error random variable does not exist, the LS method may not be the ideal estimation procedure.  
In this paper, quite counter intuitively, it is observed that both the  
least squares (LS)  and approximate LS methods provide consistent estimators of the unknown parameters of the model 
(\ref{multiple-chirp-model}).  Moreover, if the errors are from a symmetric stable 
distribution, then asymptotically LS estimators (LSEs) and approximate LS estimators (ALSEs) follow a 
multivariate symmetric stable distribution.    
 
Let us recall that a symmetric (around 0) random variable X is said to have a symmetric $\alpha$-stable ($S\alpha S$) distribution 
with the scale parameter $\sigma$ and the stability index $\alpha$, if the characteristic function of the random variable X is $E[e^{itX}] = e^{-\sigma^\alpha |t|^\alpha}$.  The $S\alpha S$ distribution is a special case of the general Stable distribution with a non-zero shift and a skewness parameters.  For different properties of the Stable and $S\alpha S$ distributions, see Samorodnitsky and Taqqu \cite{ST:1994}.  We need the following assumptions on the error process and the true values of the unknown parameters.  
\begin{assum}  \label{assum1}
The error random variables $\{e(t)\}$ is a sequence of i.i.d. random variables with mean zero and $E|e(t)|^{1+\delta} < 0$ for some $0 < \delta < 1$.
\end{assum}
\begin{assum}   \label{assum2}
The error random variables $\{e(t)\}$ is a sequence of i.i.d. random variables with mean zero and distributed as symmetric $\alpha$-stable distribution with scale parameter $\sigma$ where $1+\delta < \alpha<2$, $0 < \delta < 1$.
\end{assum}
\begin{assum} \label{assum33}
The frequencies $\theta_{1k}^0$'s and the chirp rate $\theta_{2k}^0$'s are distinct and $(A_k^0, B_k^0, \theta_{1k}^0, \theta_{2k}^0)$ is an interior point of the parameter space $\Theta$, $k=1,\ldots,p$.
\end{assum}
 \begin{assum}  \label{assum5}
The amplitudes $A_k^0$ and $B_k^0$, $k=1,\ldots,p$ satisfy the following order restrictions;
\[
\infty > M^2 \ge {A_1^0} + {B_1^0} > \cdots > {A_p^0} + {B_p^0}
\]
 \end{assum}
When $\delta \ge 1$ in Assumption \ref{assum1}, the second moment exists.  That is $\{e(t)\}$ is a sequence of i.i.d. random variables with mean zero and finite variance. 
Under Assumption \ref{assum2}, $\{e(t)\}$ is distributed as symmetric $\alpha$-stable so that $\alpha$-th moment does not exist and under Assumption \ref{assum1}, $(1+\delta)$-th moment of $\{e(t)\}$ exists.  Therefore, the condition $1+\delta < \alpha<2$, $0 < \delta < 1$ is required.  

The rest of the paper is organized as follows.  In section \ref{sec2}, we discuss the consistency properties of the LSEs and ALSEs for single chirp signal.  Asymptotic distribution of LSEs for single chirp model is established in section \ref{asym-dist-single-chirp}.  Theoretical properties for LSEs for multiple chirp signal model is derived in section \ref{asymp-multiple-chirp}.  Numerical experiments are presented in section \ref{numerical-exp} and the paper is concluded in section \ref{conclusions}.

\section{Estimation of Parameters for Single Chirp Model}  \label{sec2}
In this section, we study the asymptotic properties of LSEs and approximate LSEs for unknown parameters for chirp signal model.  We consider model \eqref{multiple-chirp-model} with $p=1$ and for notational simplicity, we write the model as 
\be
y(t) = A^0 \cos(\theta_1^0 t + \theta_2^0 t^2) + B^0 \sin(\theta_1^0 t + \theta_2^0 t^2)
+ e(t); \hspace{.15in} 
t = 1, \ldots, N.   \quad \quad \label{single-chirp-model}
\ee
Write ${\ve\xi} = (A, B, \theta_1, \theta_2)$ and ${\ve\xi}^0 = (A^0, B^0, \theta_1^0, \theta_2^0)$ as the true value of ${\ve\xi}$, then the LSE of ${\ve\xi}^0$, say $\wh{\ve\xi}$ is obtained by minimizing the following residual sum of squares;
\be
Q(\ve\xi) = \sum_{t=1}^n \Blp y(t) - A \cos(\theta_1 t + \theta_2 t^2) - B \sin(\theta_1 t + \theta_2 t^2) \Brp^2  \label{sum-square-eq}
\ee
with respect to $A$, $B$, $\theta_1$ and $\theta_2$.   We assume that the true parameter vector ${\ve\xi}^0$ is an interior point of the parameter space $\Theta$.  Minimizing \eqref{sum-square-eq} with respect to ${\ve\xi}$ requires a four dimensional optimization, but using separable regression technique of Richards \cite{Richards:1961}, it can be obtained by solving a two dimensional minimization problem in the following way. 

 Consider the following $n\times 2$ matrix
 \[
 X({\ve\theta}) = \left[ \begin{matrix}
 \cos(\theta_1  + \theta_2) & \sin (\theta_1  + \theta_2) \\
  \cos(2\theta_1  + 4\theta_2) & \sin (2\theta_1  + 4\theta_2) \\
  \vdots & \vdots \\
   \cos(n\theta_1  + n^2 \theta_2) & \sin (n \theta_1  + n^2\theta_2) \\
 \end{matrix}  \right]
 \]
 and use ${\bf a} = (A, B)^T$ and ${\bf Y} = (y(1), y(2), \ldots, y(n))^T$. Then, $$Q({\ve\xi})=({\bf Y} - X({\ve\theta}){\bf a})^T ({\bf Y} - X({\ve\theta}){\bf a}),$$ and for a given ${\ve\theta}=(\theta_1, \theta_2)$, $Q({\ve\xi})$ is minimized at 
$$\wh{\bf a}({\ve \theta}) = (X^T({\ve\theta}) X({\ve\theta}))^{-1} X^T({\ve\theta}) {\bf Y}.$$
 Therefore, replacing ${\bf a}$ by $\wh{\bf a}({\ve \theta})$ in $Q({\ve\xi})$ and minimizing $Q({\ve\xi})$ with respect to 
${\ve\xi}$ boils down to minimizing 
 \be
 Q(\wh{\bf a}({\ve \theta}), {\ve\theta}) = {\bf Y}^T({\bf I} - P_{X({\ve\theta})}){\bf Y}, \label{q_eq_separable}
\ee
with respect to ${\ve\theta}$, where
$$
P_{X({\ve\theta})} = X({\ve\theta}) (X^T({\ve\theta}) X({\ve\theta}))^{-1} X^T({\ve\theta}).
$$ 
Thus, a two step method can be implemented to obtain the LSE of ${\ve\xi}$.  First minimize $Q(\wh{\bf a}({\ve \theta}), {\ve\theta})$ with respect to ${\ve\theta}$, denote it as $\wh{\ve\theta}$ and then estimate the linear parameters $A$ and $B$ using $\wh{\bf a}(\wh{\ve \theta})$.

We propose the approximate LSEs (ALSE) of $(\theta_1, \theta_2)$ in the same line as in case of sinusoidal model, by maximizing the following periodogram like function 
\be
I({\ve\theta}) =\frac{2}{n}\left| \sum_{t=1}^n y(t) e^{- i(\theta_1 t + \theta_2 t^2)} \right|^2   \label{i-eq}
\ee
with respect to $\theta_1$ and $\theta_2$.  If $\wt{\ve\theta}=(\widetilde{\theta}_1, \widetilde{\theta}_2)$ maximizes $I({\ve\theta})$, then we call $\widetilde{\ve\theta}$ as ALSE of ${\ve\theta}$.  Following the approach of sinusoidal model, we propose the ALSEs of $A$ and $B$, say $\widetilde{A}$ and $\widetilde{B}$ as
\be  
\widetilde{A} = \frac{2}{n} \sum_{t=1}^n y(t) \cos(\widetilde{\theta}_1 t + \widetilde{\theta}_2 t^2), \;\;\;\;
\widetilde{B} = \frac{2}{n} \sum_{t=1}^n y(t) \sin(\widetilde{\theta}_1 t + \widetilde{\theta}_2 t^2).   \label{alse_eq}
\ee
Then $\wt{\ve\xi} = (\wt{A}, \wt{B}, \wt{\theta}_1, \wt{\theta}_2)$ is the ALSE ${\ve\xi}^0$.  
In case of single chirp,  the equivalent assumption of Assumption \ref{assum5} is the following. 
\begin{assum}  \label{assum6}
$A^0$ and $B^0$ are not identically equal to zero and ${A^0}^2 + {B^0}^2 < M^2 < \infty$
\end{assum}
Under assumption \ref{assum1}  on the error process $\{e(t)\}$, we prove the strong consistency of the LSEs and ALSEs and the results are stated in the following theorems.  The proofs of theorem \ref{thm1} and \ref{thm2} are provided in appendices A and B, respectively. We derive the asymptotic distribution of the LSEs and ALSEs of the unknown parameters under assumption \ref{assum2} in next section.  

\begin{theorem}  \label{thm1}
Under assumptions \ref{assum1}, \ref{assum33} and \ref{assum6}, $\wh{\ve\xi}$, the LSE of ${\ve\xi}^0$, is a strongly consistent estimator of  ${\ve\xi}^0$.
\end{theorem}
\begin{theorem}  \label{thm2}
If $A^0$ and $B^0$ are not identically equal to zero, then under assumptions \ref{assum1} and \ref{assum33}, $\wt{\ve\xi}$, the ALSE of  ${\ve\xi}^0$, defined in \eqref{i-eq} and \eqref{alse_eq}, is a strongly consistent estimator of  ${\ve\xi}^0$. 
\end{theorem}

\section{Asymptotic Distribution of LSEs and ALSEs for single chirp model}  \label{asym-dist-single-chirp} \label{sec3}

In this section, we first develop the asymptotic distribution of the LSEs of the unknown parameters for single component chirp model under Assumption \ref{assum2}. Then we discuss the equivalence of LSEs and ALSEs in their asymptotic distributions.   Under Assumption \ref{assum2}, $\{e(t)\}$ is a i.i.d. sequence of symmetric $\alpha$-stable random variables, where $1+\delta < \alpha <2$, so that $\alpha$-th moment does not exist whereas $(1+\delta)$-th moment does.   As defined earlier, $Q({\ve\xi})$ is the residual sum of squares for the single component chirp model \eqref{single-chirp-model}.  Denote $Q'({\ve \xi})$ and $Q''({\ve\xi})$ as the vector of first order derivatives of order $1\times 4$ and matrix of second order derivatives of order $4\times 4$ of $Q(\ve\xi)$, respectively.  Expand $Q'(\what{\ve \xi})$ around ${\ve \xi}^0$, using multivariate Taylor series expansion
\be
Q'(\what{\ve{\xi}}) - Q'(\ve{\xi}^0) = (\what{\ve{\xi}} -
\ve{\xi}^0) Q''(\bar{\ve{\xi}}),   \label{asym-eq-1}
\ee
where $\bar{\ve{\xi}}$ is a point on the line joining $\what{\ve{\xi}}$ and
$\ve{\xi}^0$.   Suppose ${\bf D}_1$ and ${\bf D}_2$ are two diagonal matrices defined as follows:
\beanno
{\bf D}_1 &=& \text{diag}\Blb n^{-\frac{1}{\alpha}}, n^{-\frac{1}{\alpha}}, n^{-\frac{1+\alpha}{\alpha}}, n^{-\frac{1+2\alpha}{\alpha}}
\Brb,  \\
{\bf D}_2 &=& \text{diag} \Blb n^{-\frac{\alpha-1}{\alpha}}, n^{-\frac{\alpha-1}{\alpha}}, n^{-\frac{2\alpha-1}{\alpha}}, n^{-\frac{3\alpha-1}{\alpha}} \Brb.
\eeanno
Since $\wh{\ve\xi}$ minimizes $Q({\ve\xi})$, $Q'(\what{\ve{\xi}})=0$, \eqref{asym-eq-1} can be written as 
\be
\left ( \what{\ve{\xi}} - \ve{\xi}^0 \right ) {\bf D}_2^{-1}
= - \left [ Q'(\ve{\xi}^0) {\bf D}_1 \right ] \left [ {\bf D}_2
Q''(\bar{\ve{\xi}}) {\bf D}_1 \right ]^{-1},   
\ee
provided ${\bf D}_2 Q''(\bar{\ve{\xi}}) {\bf D}_1$ is an invertible matrix for large $n$.  
It follows that $\what{\ve\xi}$ converges to ${\ve\xi}^0$ a.s. from theorem \ref{thm1} and $Q''(\ve\xi)$ is a continuous function of ${\ve\xi}$.  Therefore, using continuous mapping theorem
\be
\lim_{n \rightarrow \infty} \left [ {\bf D}_2
Q''(\bar{\ve{\xi}}) {\bf D}_1 \right ] =
\lim_{n \rightarrow \infty} \left [ {\bf D}_2
Q''(\ve{\xi}^0) {\bf D}_1 \right ] = {\bf \Gamma} \;\; \text{(say)} 
\ee
where
\be
{\bf \Gamma}=\left(\begin{matrix}
1 & 0 & \frac{B^0}{2} & \frac{B^0}{3} \\
0 & 1 & -\frac{A^0}{2} & -\frac{A^0}{3} \\
\frac{B^0}{2} & -\frac{A^0}{2} & \frac{1}{3}({A^0}^2 + {B^0}^2) & \frac{1}{4}({A^0}^2 + {B^0}^2) \\
\frac{B^0}{3} & -\frac{A^0}{3} & \frac{1}{4}({A^0}^2 + {B^0}^2) & \frac{1}{5}({A^0}^2 + {B^0}^2) \\
\end{matrix} \right)  \label{gamma_eq}
\ee
and 
\be
{\bf \Gamma}^{-1}= ((\gamma^{ij}))=\frac{1}{{A^0}^2 + {B^0}^2}\left(\begin{matrix}
{A^0}^2 + 9{B^0}^2 & -8A^0 B^0 & 36 B^0 & -30 B^0 \\
-8A^0 B^0 & 9{A^0}^2 + {B^0}^2 & -36 A^0 & 30 A^0 \\
36 B^0 & -36 A^0 & 192 & 180 \\
-30 B^0 & 30 A^0 & 180 & 180 \\
\end{matrix} \right).
\ee
In order to show that $Q'(\ve{\xi}^0) {\bf D}_1$ converges to a multivariate stable distribution, we write $Q'(\ve{\xi}^0) {\bf D}_1=(Z_{1n}, Z_{2n}, Z_{3n}, Z_{4n})$ where 
\beanno
Z_{1n} &=& -\frac{2}{n^{\frac{1}{\alpha}}} \sum_{t=1}^n e(t) \cos(\theta_1^0 t + \theta_2^0 t^2), \;\;\;\;
Z_{2n} = -\frac{2}{n^{\frac{1}{\alpha}}} \sum_{t=1}^n e(t) \sin(\theta_1^0 t + \theta_2^0 t^2), \\
Z_{3n} &=& \frac{2}{n^{\frac{1+\alpha}{\alpha}}} \sum_{t=1}^n e(t) t g({\ve \xi}^0;t), \;\;\; \;\;\;\;\;\;\;\;\;\;\;\;
Z_{4n} = \frac{2}{n^{\frac{1+2\alpha}{\alpha}}} \sum_{t=1}^n e(t) t^2 g({\ve \xi}^0;t), 
\eeanno
and $g({\ve \xi}^0;t) = A^0 \sin(\theta_1^0 t + \theta_2^0 t^2) - B^0 \cos(\theta_1^0 t + \theta_2^0 t^2)$.   If ${\bf t} = (t_1, t_2, t_3, t_4)$, the joint characteristic function of  $Q'(\ve{\xi}^0) {\bf D}_1$ is 
\beanno
\phi_n({\bf t}) = E\bls \exp\{i(t_1 Z_{1n}+t_2 Z_{2n}+t_3 Z_{3n}+t_4 Z_{4n})\} \brs 
 = E\bls \exp \blb i \frac{2}{n^{\frac{1}{\alpha}}} \sum_{r=1}^n e(r) K_{\bf t}(r) \brb \brs,
\eeanno
where 
$$
K_{\bf t}(r;{\ve\xi}^0) = -t_1 \cos(\theta_1^0 r + \theta_2^0 r^2) -t_2 \sin(\theta_1^0 r + \theta_2^0 r^2) + \frac{rt_3}{n} g({\ve\xi}^0;r) + \frac{r^2 t_4}{n^2} g({\ve\xi}^0;r).
$$
Because $\{e(t)\}$ is a sequence of independent random variables,
\beanno
\phi_n({\bf t})  = \prod_{r=1}^n E \exp \Blb i \frac{2}{n^{\frac{1}{\alpha}}} e(r) K_{\bf t}(r;{\ve\xi}^0) \Brb &=& \prod_{r=1}^n \exp \Blb -2^\alpha \sigma^\alpha \frac{1}{n} | K_{\bf t}(r; {\ve\xi}^0)|^\alpha \Brb \\
&=& \exp \Blb -2^\alpha \sigma^\alpha \frac{1}{n} \sum_{r=1}^n | K_{\bf t}(r; {\ve\xi}^0)|^\alpha \Brb.
\eeanno
It is noted similarly as in Nandi et al. \cite{NIK:2002} that $\frac{1}{n} \sum_{r=1}^n | K_{\bf t}(r; {\ve\xi}^0)|^\alpha$ converges as 
$n \rightarrow \infty$.  Now the following arguments show that it converges to a non-zero limit for ${\bf t} \ne {\bf 0}$.  
First, observe that for all $r$ and $n$; $1\le r \le n$, $n=1, 2, \ldots$
\[
| K_{\bf t}(r; {\ve\xi}^0)| \le |t_1| + |t_2| + (|t_3|+|t_4|)(A^0+B^0) = S, \;\;\; \text{(say)}.
\]
Therefore, $| K_{\bf t}(r; {\ve\xi}^0)/S| \le 1$.  Hence, for $0 < \alpha \le 2$ and $r=1,2,\ldots$
\[
| K_{\bf t}(r; {\ve\xi}^0)/S|^\alpha  \ge  | K_{\bf t}(r; {\ve\xi}^0)/S|^2  \Rightarrow | K_{\bf t}(r; {\ve\xi}^0)|^\alpha  \ge \frac{S^\alpha}{S^2}  |K_{\bf t}(r; {\ve\xi}^0)|^2.
\] 
Using Lemma \ref{lemma0}(stated in Appendix A), it follows that 
\[
\lim_{n \rightarrow\infty}\frac{1}{n} \sum_{r=1}^n | K_{\bf t}(r; {\ve\xi}^0)|^\alpha = \tau_{\bf t}(A^0, B^0, \theta_1^0, \theta_2^0, \alpha) = \tau_{\bf t}({\ve\xi}^0;\alpha) > 0.
\]
This implies that 
\be
\lim_{n\rightarrow \infty} \phi_n({\bf t}) = \exp \Blb -2^\alpha \sigma^\alpha  \tau_{\bf t}({\ve\xi}^0;\alpha)  \label{asym-eq-2}
\Brb.
\ee
This limiting characteristic function \eqref{asym-eq-2} implies that any linear combination of $Z_{1n}$, $Z_{2n}$, $Z_{3n}$ and  $Z_{4n}$, even for large samples, follows a symmetric stable distribution with stability index $\alpha$.

Now, consider the following for large $n$,
\be   \label{asym-eq-3}
\left [ Q'(\ve{\xi}^0) {\bf D}_1 \right ] \left [ {\bf D}_2
Q''(\bar{\ve{\xi}}) {\bf D}_1 \right ]^{-1}  = \left[\begin{matrix}
\sum_{k=1}^4 Z_{kn} \gamma^{k1} \\
\sum_{k=1}^4 Z_{kn} \gamma^{k2} \\
\sum_{k=1}^4 Z_{kn} \gamma^{k3} \\
\sum_{k=1}^4 Z_{kn} \gamma^{k4} 
\end{matrix}  \right]
= -\frac{2}{n^{\frac{1}{\alpha}}} \left[\begin{matrix}
\sum_{r=1}^n e(r) U_{A}(r;{\ve\xi}^0) \\
\sum_{r=1}^n e(r) U_{B}(r;{\ve\xi}^0) \\
\sum_{r=1}^n e(r) U_{\theta_1}(r;{\ve\xi}^0) \\
\sum_{r=1}^n e(r) U_{\theta_2}(r;{\ve\xi}^0) 
\end{matrix} \right]
\ee
where
\beanno
U_{A}(r;{\ve\xi}) &=& \frac{1}{{A}^2+{B}^2}  \Blp  ({A}^2 + 9 {B}^2)\cos(\theta_1 r + \theta_2 r^2) -8 A B \sin(\theta_1 r + \theta_2 r^2) \\
& & \hspace{1in}- \frac{36B r}{n} g({\ve\xi};r) + \frac{30 B r^2}{n^2} g({\ve\xi};r) \Brp,  \\
U_{B}(r;{\ve\xi}) &=& \frac{1}{{A}^2+{B}^2}  \Blp  -8 A B \cos(\theta_1 r + \theta_2 r^2) + (9{A}^2 + {B}^2)\sin(\theta_1 r + \theta_2 r^2) \\
& & \hspace{1in} + \frac{36A r}{n} g({\ve\xi};r) - \frac{30 A r^2}{n^2} g({\ve\xi};r) \Brp,  \\
U_{\theta_1}(r;{\ve\xi}) &=& \frac{1}{{A}^2+{B}^2}  \Blp  36B \cos(\theta_1 r + \theta_2 r^2) -36 A \sin(\theta_1 r + \theta_2 r^2) \\
& & \hspace{1in} - \frac{192 r}{n} g({\ve\xi};r) - \frac{180 r^2}{n^2} g({\ve\xi};r) \Brp,  \\
U_{\theta_2}(r;{\ve\xi}) &=& \frac{1}{{A}^2+{B}^2}  \Blp  -30 B \cos(\theta_1 r + \theta_2 r^2) + 30 A \sin(\theta_1 r + \theta_2 r^2) \\
& & \hspace{1in} - \frac{180 r}{n} g({\ve\xi};r) - \frac{180 r^2}{n^2} g({\ve\xi};r) \Brp.  \\
\eeanno
Each element of $\left [ Q'(\ve{\xi}^0) {\bf D}_1 \right ] \left [ {\bf D}_2
Q''(\bar{\ve{\xi}}) {\bf D}_1 \right ]^{-1} $ is a linear combination of $Z_{1n}$, $Z_{2n}$, $Z_{3n}$ and  $Z_{4n}$ and therefore, distributed as symmetric $\alpha$-stable distribution.  Using Theorem 2.1.5 of Samorodnitsky and Taqqu \cite{ST:1994} that {\it a
random vector is
symmetric $\alpha$-stable in ${\mathbb R}^d$, if and only if any linear combination is symmetric stable in ${\mathbb R}^1$, where $d$ is the order of the
vector}, it immediately follows that
\[
\left [ Q'(\ve{\xi}^0) {\bf D}_1 \right ] \left [ {\bf D}_2
Q''(\bar{\ve{\xi}}) {\bf D}_1 \right ]^{-1}
\]
converges to a symmetric $\alpha$-stable random vector in $\mathbb{R}^4$ as $n\rightarrow \infty$ and has the following limiting characteristic function 
\be
\phi({\bf t}) = \exp \Blb -2^\alpha \sigma^\alpha  \tau_{\bf v}({\ve\xi}^0;\alpha)  \Brb.  \label{asym-eq-4}
\ee
The vector ${\bf v}$ is defined as in \eqref{asym-eq-2}, replacing ${\bf t}$ by ${\bf v}$ where 
\beanno
v_1({\bf t}; A^0, B^0) &=& \frac{1}{{A^0}^2+{B^0}^2}  \Blp ({A^0}^2 + 9 {B^0}^2)t_1 -8A^0 B^0 t_2 + 36 B^0 t_3 - 30 B^0 t_4 \Brp,  \\
v_2({\bf t}; A^0, B^0) &=& \frac{1}{{A^0}^2+{B^0}^2}  \Blp -8A^0 B^0 t_1 + (9{A^0}^2 + {B^0}^2)t_2 -36 B^0 t_3 + 30 A^0 t_4 \Brp, \\
v_3({\bf t}; A^0, B^0) &=& \frac{1}{{A^0}^2+{B^0}^2}  \Blp 36 B^0 t_1 - 36 A^0 t_2 +192 t_3 + 180 t_4 \Brp, \\
v_4({\bf t}; A^0, B^0) &=& \frac{1}{{A^0}^2+{B^0}^2}  \Blp -30 B^0 t_1 + 30 A^0 t_2 +180 t_3 + 180 t_4 \Brp. 
\eeanno
So, we  have the following theorem.
\begin{theorem}  \label{thm3}
In case of model \eqref{single-chirp-model}, under Assumptions \ref{assum2}, \ref{assum33}  and \ref{assum6}, 
\[
(\wh{\ve\xi} - \ve\xi^0){\bf D}_2^{-1} = \Blp n^{\frac{\alpha - 1}{\alpha}} (\widehat{A} - A^0), n^{\frac{\alpha -1}{\alpha}} (\widehat{B} - B^0), n^{\frac{2\alpha - 1}{\alpha}}  (\widehat{\theta}_1 - \theta_1^0),
n^{\frac{3\alpha - 1}{\alpha}} (\widehat{\theta}_2 - \theta_2^0) \Brp   
\]
converges to a multivariate $S\alpha S$ distribution in $\mathbb{R}^4$ having the characteristic function given in \eqref{asym-eq-4}.
\end{theorem}
It has been observed that ALSEs have the same asymptotic distribution as the LSEs in case of model \ref{single-chirp-model} and the result is stated in the following theorem.  

\begin{theorem}  \label{thm33}
If $A^0$ and $B^0$ are not identically equal to zero, then under Assumptions \ref{assum2} and \ref{assum33}, the limiting distribution of 
\[
(\wt{\ve\xi} - \ve\xi^0){\bf D}_2^{-1}= \Blp n^{\frac{\alpha - 1}{\alpha}} (\wt{A} - A^0), n^{\frac{\alpha -1}{\alpha}} (\wt{B} - B^0), n^{\frac{2\alpha - 1}{\alpha}}  (\wt{\theta}_1 - \theta_1^0),
n^{\frac{3\alpha - 1}{\alpha}} (\wt{\theta}_2 - \theta_2^0) \Brp   
\]
is same as the limiting distribution of $(\wh{\ve\xi} - \ve\xi^0){\bf D}_2^{-1}$.
\end{theorem}

\noindent {\bf Proof of Theorem \ref{thm33}:} See in Appendix C.

\section{Consistency and asymptotic distribution for multiple chirp model}  \label{asymp-multiple-chirp} 

In this section, we discuss the asymptotic results of the LSEs of the unknown parameters for multiple chirp model \eqref{multiple-chirp-model}.  The parameter vector is denoted by ${\eta} = ({\ve\eta}_1, \ldots, {\ve\eta}_p)$, ${\ve \eta}_k=(A_k, B_k, \theta_{1k}, \theta_{2k})$, $k=1,\ldots,p$ and ${\ve\eta}^0$  denote the true value of ${\ve\eta}$.   The LSE of ${\ve\eta}$, say $\wh{\ve\eta}$ is obtained by minimizing the residual sum of squares for model \eqref{multiple-chirp-model}, defined similarly as $Q(\ve\xi)$ in \eqref{sum-square-eq}.  We write $R(\ve\eta)$ as the residual sum of squares in this case and defined as follows;
\be
R(\ve\eta) = \sum_{t=1}^n \left[y(t) - \sum_{k=1}^p \Blp A_k \cos(\theta_{1k} t + \theta_{2k} t^2) + B_k \sin(\theta_{1k} t + \theta_{2k} t^2) \Brp \right]^2.
\ee
In matrix notation, 
\beanno
R(\ve\eta) & = & \left ({\bf Y} - X_p({\ve\theta}){\ve\psi} \right )^T \left ({\bf Y} - X_p({\ve\theta}){\ve\psi} \right )  \\
& = & \left ({\bf Y} - \sum_{k=1}^p X({\ve\theta}_k){\bf a}_k \right )^T \left ({\bf Y} - \sum_{k=1}^p X({\ve\theta}_k ){\bf a}_k \right ), 
\eeanno
where $X_p({\ve\theta}) = [X({\ve\theta}_1), \ldots, X({\ve\theta}_p)]$ and $X({\ve\theta}_k)$ is same as $X({\ve\theta})$, ${\ve\theta}$ replaced by ${\ve\theta}_k = (\theta_{1k}, \theta_{2k})$; ${\ve\psi}^T = ({\bf a}_1, \ldots, {\bf a}_p)$, ${\bf a}_k = (A_k, B_k)^T$. Then, for a given ${\ve\theta}=({\ve\theta}_1, \ldots, {\ve\theta}_p)$, the vector of linear parameters ${\ve\psi}$ can be estimated as 
$$
\wh{\ve\psi}({\ve\theta}_1, \ldots, {\ve\theta}_p) =  (X_p^T({\ve\theta}) X_p({\ve\theta}))^{-1} X_p^T({\ve\theta}) {\bf Y}. 
$$
Replacing $\wh{\ve\psi}({\ve\theta}_1, \ldots, {\ve\theta}_p)$ in $R({\ve\eta})$, similarly as single chirp model, ${\theta}_{1k}$ and ${\ve\theta}_{2k}$, $k=1,\ldots, p$ can be estimated. Finally, ${\ve\psi}$ is estimated as $\wh{\ve\psi}(\wh{\ve\theta}_1, \ldots, \wh{\ve\theta}_p)$.

The ALSEs of the unknown parameters in case of model \eqref{multiple-chirp-model} are obtained similarly as the single chirp model.  We maximize the periodogram like function
\be
I({\ve\theta}) =\frac{2}{n}\left| \sum_{t=1}^n y(t) e^{- i(\theta_1 t + \theta_2 t^2)} \right|^2   \label{i-eq-multiple-chirp}
\ee
with respect to ${\theta}_1$ and ${\theta}_2$, locally and sequentially.  Under assumption \ref{assum5}, at the first step, $(\theta_1,\theta_2)$ that maximizes $I({\ve\theta})$ is the estimate of $(\theta_{11}, \theta_{21})$; write it as  $(\wt{\theta}_{11}, \wt{\theta}_{21})$.  At the second step, we obtain the estimate of $(\theta_{11}, \theta_{21})$, say $(\wt{\theta}_{21}, \wt{\theta}_{22})$ and at the $p$-th step we the estimate of $(\theta_{p1}, \theta_{p1})$, say $(\wt{\theta}_{p1}, \wt{\theta}_{p2})$.  The linear parameters $A_k$ and $B_k$ are estimated as 
\be
\widetilde{A}_k = \frac{2}{n} \sum_{t=1}^n y(t) \cos(\widetilde{\theta}_{1k} t + \widetilde{\theta}_{2k} t^2), \;\;\;\;
\widetilde{B} = \frac{2}{n} \sum_{t=1}^n y(t) \sin(\widetilde{\theta}_{1k} t + \widetilde{\theta}_{2k} t^2), \;\; k=1,\ldots,p.  \label{alse-linear-multiple}
\ee
We write $\wt{\ve\eta}_k$  as the ALSE of ${\ve\eta}_k$.  We also note that the boundedness condition on linear parameters (stated in Assumption \ref{assum5}) is not required in case of ALSEs.  The LSEs and ALSEs of the unknown parameters in case of multiple chirp model \eqref{multiple-chirp-model}, similarly as single chirp model \eqref{single-chirp-model}, provide strongly consistent estimators.    We state the results in the following theorems.
\begin{theorem}  \label{thm4}
In model \eqref{multiple-chirp-model}, if the error process $\{e(t)\}$ satisfies Assumption \ref{assum1}, frequencies satisfy Assumption \ref{assum33} and  amplitudes $A_j^0$ and $B_j^0$, $j=1,\ldots,p$
 satisfy Assumption \ref{assum5}, then $\wh{\ve\eta}_j$ is a strongly consistent estimator of ${\ve\eta}_j^0$, $j=1,\ldots,p$.
\end{theorem}
\begin{theorem}  \label{thm5}
In model \eqref{multiple-chirp-model}, if the error process $\{e(t)\}$ satisfies Assumption \ref{assum1}, frequencies satisfy Assumption \ref{assum33} and amplitudes $A_j^0$ and $B_j^0$, $j=1,\ldots,p$ satisfy
$$
\infty > {A_1^0} + {B_1^0} > \cdots > {A_p^0} + {B_p^0}
$$
 then $\wt{\ve\eta}_j$ is a strongly consistent estimator of ${\ve\eta}_j^0$, $j=1,\ldots,p$.
\end{theorem}
Theorems \ref{thm4} and \ref{thm5} can be proved along the same lines as theorems \ref{thm1} and \ref{thm2}, so the proof are not provided.

We discuss the asymptotic distribution of the LSEs of the unknown parameters, $\wh{\ve\eta}$ under Assumptions \ref{assum2}, \ref{assum33} and \ref{assum5}.  Similarly as previous section, let $R'({\ve\eta})$ and $R''({\ve\eta})$ denote the vector of first derivatives and matrix of second derivatives of $R({\ve\eta})$ of orders $1\times 4p$ and $4p \times 4p$, respectively.  Define two diagonal matrices ${\ve\Delta}_1$ and ${\ve\Delta}_2$ of order $4p \times 4p$ as follows:
\[
{\ve\Delta}_1 = \left[\begin{matrix}
{\bf D}_1 & {\bf 0} &  \cdots & {\bf 0} \\
{\bf 0} & {\bf D}_1 & \cdots & {\bf 0} \\
\vdots & \vdots & \vdots & \vdots \\
{\bf 0} &  {\bf 0} & \cdots & {\bf D}_1
\end{matrix} \right],  \;\;\;\; \;\;
{\ve\Delta}_2 = \left[\begin{matrix}
{\bf D}_2 & {\bf 0} &  \cdots & {\bf 0} \\
{\bf 0} & {\bf D}_2 & \cdots & {\bf 0} \\
\vdots & \vdots & \vdots & \vdots \\
{\bf 0} &  {\bf 0} & \cdots & {\bf D}_2
\end{matrix} \right],
\]
where ${\bf D}_1$ and ${\bf D}_2$ are same as defined in section \ref{asym-dist-single-chirp}.  Using multivariate Taylor series expansion along the same line as in section \ref{asym-dist-single-chirp}, we have 
\be
\left ( \what{\ve{\eta}} - \ve{\eta}^0 \right ) {\ve\Delta}_2^{-1}
= - \left [ R'(\ve{\eta}^0) {\ve\Delta}_1 \right ] \left [ {\ve\Delta}_2
R''(\bar{\ve{\eta}}) {\ve\Delta}_1 \right ]^{-1}, 
\ee
because $R'(\wh{\ve\eta}) ={\bf 0}$ and  ${\ve\Delta}_2
R''(\bar{\ve\eta}) {\ve\Delta}_1$ is an invertible matrix for large $n$.  Here, $\bar{\ve{\eta}}$ is a point on the line joining $\wh{\ve\eta}$ and ${\ve\eta}^0$.  Similarly, as in case of model \eqref{single-chirp-model} for ${\bf D}_2 Q''(\bar{\ve\xi}){\bf D}_1$, we can show that 
\[
\lim_{n\rightarrow\infty} \left [ {\ve\Delta}_2
R''(\bar{\ve{\eta}}) {\ve\Delta}_1 \right] = \lim_{n\rightarrow\infty} \left [ {\ve\Delta}_2
R''({\ve\eta}^0) {\ve\Delta}_1 \right] = \left[\begin{matrix}
{\ve\Gamma}_1 & {\bf 0} &  \cdots & {\bf 0} \\
{\bf 0} & {\ve\Gamma}_1 & \cdots & {\bf 0} \\
\vdots & \vdots & \vdots & \vdots \\
{\bf 0} &  {\bf 0} & \cdots & {\ve\Gamma}_1
\end{matrix} \right] = {\ve\Gamma}_G, \;\;\; \text{(say)},
\]
where ${\ve \Gamma}_k$ is a $4 \times 4$ matrix obtained from ${\ve \Gamma}$ by replacing $A^0$ and $B^0$ by $A_k^0$ and
$B_k^0$, respectively.   Consider ${\bf t} = ({\bf t_1}, \ldots, {\bf t_p})$, ${\bf t_j} = (t_{1j},  t_{2j}, t_{3j},
t_{4j})$ and write $R'({\ve \eta}){\ve \Delta}_2 = ({\bf Z}_n^1, \ldots, {\bf Z}_n^p)$, ${\bf Z}_n^j = (Z_{1n}^{j},
Z_{2n}^{j},Z_{3n}^{j},Z_{4n}^{j})$.  The elements of ${\bf Z}_n^j$, $Z_{kn}^{j}$, $k=1,\ldots,4$ are defined similarly as $Z_{kn}$, $k=1,\ldots,4$;  $A^0$, $B^0$, $\theta_1^0$ and $\theta_2^0$ replaced by $A_k^0$, $B_k^0$, $\theta_{1k}^0$ and $\theta_{2k}^0$, respectively.  With these notation, the joint characteristic function of the elements of $R'({\ve \eta}){\ve \Delta}_2$ is 
\beanno
\phi_n^p({\bf t}) &=& E \exp \Blb i \frac{2}{n^{\frac{1}{\alpha}}} \sum_{r=1}^n e(r) K_{\bf t}^M(r;{\ve\eta}^0)  \Brb, \;\;\;   K_{\bf t}^M(r;{\ve\eta}^0) = \sum_{k=1}^p K_{\bf t_k}(r;{\ve\eta}_k^0).
\eeanno
$K_{\bf t_k}(r;{\ve\eta}_k^0)$ is same as $K_{\bf t}(r;{\ve\xi}^0)$ with $A^0$, $B^0$, $\theta_1^0$ and $\theta_2^0$ replaced by $A_k^0$, $B_k^0$, $\theta_{1k}^0$ and $\theta_{2k}^0$, respectively for $k=1,\ldots,p$.  This particular form of $K_{\bf t}^M(r;{\ve\eta}^0)$ and independence of $\{e(r)\}$ enable us to write
\beanno
\phi_n^p({\bf t}) = \prod_{j=1}^p \prod_{r=1}^n  \exp \Blb i \frac{2}{n^{\frac{1}{\alpha}}}  e(r) K_{\bf t_k}(r;{\ve\eta}_k^0) \Brb 
&=& \prod_{j=1}^p \prod_{r=1}^n  \exp \Blb - \frac{2^\alpha \sigma^\alpha}{n} |K_{\bf t_k}(r;{\ve\eta}_k^0)|^\alpha \Brb \\
&=& \prod_{j=1}^p \exp \Blb  -\frac{2^\alpha \sigma^\alpha}{n} \sum_{r=1}^n  |K_{\bf t_k}(r;{\ve\eta}_k^0)|^\alpha \Brb.
\eeanno
Take limit as $n\rightarrow \infty$, 
\beanno
\lim_{n\rightarrow \infty} \phi_n^p({\bf t}) &=& \prod_{j=1}^p \exp \Blb -2^\alpha \sigma^\alpha \tau_{\bf t_j}({\ve\eta}_j^0, \alpha) \Brb  \\
& = &\prod_{j=1}^p \blb \text{Joint characteristic function of}  (Z_{1n}^{j},
Z_{2n}^{j},Z_{3n}^{j},Z_{4n}^{j}) \brb.
\eeanno
This implies that $(Z_{1n}^{j}, Z_{2n}^{j}, Z_{3n}^{j}, Z_{4n}^{j})$ and $(Z_{1n}^{k}, Z_{2n}^{k}, Z_{3n}^{k},Z_{4n}^{k})$  for $j\ne k = 1,\ldots, p$ are asymptotically independently distributed.  

Now considering linear combinations similarly as in section \ref{asym-dist-single-chirp},  it is observed that as $n \rightarrow \infty$, 
$$(\wh{\ve \eta} - {\ve \eta}^0) {\ve \Delta}_2^{-1} = \Blp (\wh{\ve \eta}_1 - {\ve \eta}_1^0) {\bf D_2}^{-1}, \ldots,
(\wh{\ve
\eta}_p - {\ve \eta}_p^0) {\bf D_2}^{-1} \Brp$$ converges to a $S\alpha S$ random vector in $\mathbb{R}^{4p}$ having the
following
characteristic function 
\be 
{\ve \phi}_{\bf t}^p = \exp \Blb - 2^\alpha \sigma^\alpha \sum_{j=1}^p \tau_{\bf w_j} ({\ve\eta}_j^0, \alpha) \Brb, 
\ee
\be
{\bf w_j} = (w_{1j}({\bf t_j}),  w_{2j}({\bf t_j}),w_{3j}({\bf t_j}),w_{4j}({\bf t_j})),   \;\;w_{kj}({\bf t_j}) =  v_k ({\bf t_j}; A_j^0, B_j^0), \;\; k=1,\ldots,4; \;\; 
j=1,\ldots, p,
\label{w-eq}
\ee
where $v_k$, $k=1,\ldots,4$ are defined in section \ref{asym-dist-single-chirp}.  Therefore, we state the asymptotic distribution of the LSEs of the unknown parameters in model \eqref{multiple-chirp-model} in the following theorem.
\begin{theorem}  \label{thm6}
In model \eqref{multiple-chirp-model}, if the error process $\{e(t)\}$ satisfies Assumption \ref{assum2}, frequencies satisfy Assumption \ref{assum33} and  amplitudes $A_j^0$ and $B_j^0$, $j=1,\ldots,p$
 satisfy Assumption \ref{assum5}, then $(\wh{\ve\eta}_j - {\ve\eta}_j^0) {\bf D}_2^{-1}$ converges to a multivariate $\alpha$-stable distribution in $\mathbb{R}^4$ with limiting characteristic function $\exp\{-2^\alpha \sigma^\alpha \tau_{\bf w_j} ({\ve\eta}_j^0, \alpha) \}$ where ${\bf w_j}$ is defined in \eqref{w-eq}.  Also $(\wh{\ve\eta}_j - {\ve\eta}_j^0) {\bf D}_2^{-1}$ and $(\wh{\ve\eta}_k - {\ve\eta}_k^0) {\bf D}_2^{-1}$ for $j\ne k$ are asymptotically independently distributed. 
\end{theorem}
In case of multiple chirp model, ALSEs are asymptotically equivalent to LSEs in distribution and it can be shown similarly as Theorem \ref{thm33}.
The joint asymptotic distribution of the ALSEs, $\wt{\ve\eta}_1, \ldots, \wt{\ve\eta}_p$ is same as that of the $\wh{\ve\eta}_1, \ldots, \wh{\ve\eta}_p$ and stated in the following theorem.
\begin{theorem}  \label{thm7}
Under Assumptions \ref{assum2}, \ref{assum33}, if  model \eqref{multiple-chirp-model} satisfies 
$$
\infty >  {A_1^0} + {B_1^0} > \cdots > {A_p^0} + {B_p^0},
$$
then the limiting distribution of $(\wt{\ve\eta}_j - {\ve\eta}_j^0){\bf D}_2^{-1}$ 
is same as the limiting distribution of $(\wh{\ve\eta}_ - {\ve\eta}_j^0){\bf D}_2^{-1}$ and $(\wt{\ve\eta}_j - {\ve\eta}_j^0) {\bf D}_2^{-1}$ and $(\wt{\ve\eta}_k - {\ve\eta}_k^0) {\bf D}_2^{-1}$ for $j\ne k$ are asymptotically independently distributed.
\end{theorem}

 \section{Numerical Experiments}  \label{numerical-exp}
 
 In this section, we present results of numerical experiments.  We consider the following two models, first one is a single chirp signal model and second one is a two component multiple chirp model. 
 \bea
&& \text{Model}\; 1:\;\;y(t) = A\cos(\theta_1^0 t + \theta_2^0 t^2) + B \sin(\theta_1^0 t + \theta_2^0 t^2) + e(t), t=1,\ldots,n \label{model1_numerical} \\
&& \text{Model}\; 2: \;\; y(t) = A_1\cos(\theta_{11}^0 t + \theta_{12}^0 t^2) + B_1 \sin(\theta_{11}^0 t + \theta_{12}^0 t^2) + \nonumber \\
&&     \hspace{1.in}  A_2\cos(\theta_{21}^0 t + \theta_{22}^0 t^2) + B_2 \sin(\theta_{21}^0 t + \theta_{22}^0 t^2) + e(t), t=1,\ldots,n   \label{model2_numerical}
\eea
$$A^0=B^0=2.5, \;\; \theta_1^0 = 1.5 \;\;\text{and} \;\;\theta_2^0 =0.1,$$
$$A_1^0 = B_1^0 = 4.0, \;\; \theta_{11}^0=1.5, \;\; \theta_{12}^0 = .1 \;\; A_2^0 = B_2^0 = 3.0, \;\; \theta_{11}^0=2.5, \;\; \theta_{12}^0 = 0.2.$$
 The sequence of random variables $\{e(t)\}$ is from a symmetric stable distribution with mean zero, scale parameters $\sigma>0$ and stability index $\alpha < 2$.  We have taken the value of the frequency rate to
be much less than the initial frequency, as the frequency rate represents the rate of change and is comparatively small in general.  We would like to see how the LSEs and ALSEs of the unknown parameters, specially the nonlinear parameters,  behave for different values of $\alpha$, $\sigma$ and $n$, the sample size.  That is, for single chirp model \eqref{model1_numerical},  the frequency $\theta_1$ and the frequency rate $\theta_2$ and for model \eqref{model2_numerical}, the frequencies $\theta_{11}$ and $\theta_{21}$  and frequency rates $\theta_{12}$ and $\theta_{22}$.  For simulation, we consider $\alpha=1.5, 1.7, 1.9$; $\sigma=0.1, 1.0$ and $n=250, 500, 1000$.  
 
We generate a sample of size $n$ for each combination of $\alpha$, $\sigma$ and $n$ and compute the LSEs and ALSEs of the unknown parameters.  We replicate the process of data generation and estimation $5000$ times and calculate the average estimate (AVE) and mean absolute deviation (MAD) for each parameter estimator over these replications.  We have not considered mean squared error as a criterion for evaluating the estimators because in the present set-up, estimators are asymptotically multivariate stable such that the variance does not exist for $\alpha<2$.   The results for single chirp model \eqref{model1_numerical} are reported in Tables \ref{table1}-\ref{table3} for sample sizes $250$, $500$ and $1000$, respectively.  The results for two component model \eqref{model2_numerical} are presented in Tables \ref{table4}-\ref{table6}.  It is worth to be noted that for obtaining LSEs, we need to implement a four dimensional minimization whereas for ALSEs a two dimensional maximization is required for single chirp model.  For model \eqref{model2_numerical}, as the number of chirp component $p=2$, the LSEs are obtained by solving a $4p =8$ dimensional minimization problem and ALSEs are obtained by solving $p=2$, two dimensional maximization problem.  Therefore, as $p$ increases, the complexity in finding LSEs is much more as compared to the ALSEs.  We have used downhill simplex method to carry out these optimizations. 
 
 Some of the salient features of the numerical experiments based on Tables \ref{table1}-\ref{table6} are given below.
 \begin{enumerate}
 \item Performances of both LSEs and ALSEs improve as sample size $n$ increases, scale parameters $\sigma$ decreases and the stability index $\alpha$ increases.
 \item The estimates are close to true values in all the cases considered as biases are quite small in both LSEs and ALSEs in general.  Though, biases are small, ALSEs have slightly larger bias in case of frequency estimates.
 \item As expected, mean absolute deviation decreases as sample size increases .  The rate of convergence $O(\frac{1}{n^{\frac{3}{2}}})$ for frequencies and  $O(\frac{1}{n^{\frac{5}{2}}})$ for frequency rates are in general observed in each mean absolute deviation values in all the cases for LSEs and in most of the cases for ALSEs. 
 \item Mean absolute deviation increases as the scale parameter, $\sigma$ increases.  
\item  Mean absolute deviation decreases as the stability index $\alpha$ increases. It indicates that for heavier tail, it is more difficult to estimate the unknown parameters. 
\item Comparing LSE and ALSE, it is observed that although they are asymptotically equivalent, LSE behaves marginally better than the ALSE, in terms of minimum mean absolute deviation for most of the cases considered here.  
\item Computationally, ALSEs are much easier to compute than the LSEs, at least for large $p$.  Therefore, if $p$ is large, ALSE 
is preferable, but when $p$ is small, LSE is recommended.  
 \end{enumerate}
\begin{table}
\caption{Average estimates and mean absolute deviations of the estimators of frequency and frequency rate for Model 1 when the sample size $n=250$.}  \label{table1}
\begin{tabular}{|c|c|c|c|c|c|c|}
\hline
$\sigma$ & $\alpha$ &  & \multicolumn{2}{|c|}{LSE} & \multicolumn{2}{|c|}{ALSE} \\
\cline{4-7}
&  &   & $\theta_1$ & $\theta_2$  & $\theta_1$ & $\theta_2$\\
\hline 
.1 & \multirow{2}{*}{$1.5$} & AVE       &  1.50001      &    9.9997e-2   &    1.49579      &   .10002  \\
\cline{3-7}
& & MAD & 1.6988e-4  &    8.5680e-7  &    4.2101e-3  &    1.3754e-5 \\
\cline{2-7}
& \multirow{2}{*}{$1.7$} & AVE  &     1.50001      &    9.9996e-2 &    1.49576      &   .10002   \\
\cline{3-7}
& & MAD &   1.0927e-4  &    5.5196e-7 &    4.2383e-3  &    1.3843e-5  \\
\cline{2-7}
& \multirow{2}{*}{$1.9$} & AVE  &  1.50002      &    9.9996e-2  &    1.49574      &   .10002   \\
\cline{3-7}
& & MAD & 7.9631e-5  &    4.0493e-7     &    4.2544e-3  &    1.3894e-5  \\
\cline{1-7}
1.0 & \multirow{2}{*}{$1.5$} & AVE       & 1.50001      &    9.9999e-2 &    1.49605       &   .10001     \\
\cline{3-7}
& & MAD &  1.7665e-3  &    8.9748e-6  &    5.5351e-3  &    2.0923e-5 \\
\cline{2-7}
& \multirow{2}{*}{$1.7$} & AVE  &   1.50001      &    9.9998e-2 &    1.49596      &   .10001   \\
\cline{3-7}
& & MAD &  1.0913e-3  &    5.5570e-6  &    4.4925e-3  &    1.5973e-5   \\
\cline{2-7}
& \multirow{2}{*}{$1.9$} & AVE  &  1.50000     &    9.9997e-2 &  1.49592      &   .10001   \\
\cline{3-7}
& & MAD &   7.8374e-4  &    4.0271e-6  &    4.1747e-3  &    1.4240e-5   \\
\cline{1-7}
\end{tabular}
\end{table}

\begin{table}
\caption{Average estimates and mean absolute deviations of the estimators of frequency and frequency rate for Model 1 when the sample size $n=500$.}  \label{table2}
\begin{tabular}{|c|c|c|c|c|c|c|}
\hline
$\sigma$ & $\alpha$ &  & \multicolumn{2}{|c|}{LSE} & \multicolumn{2}{|c|}{ALSE} \\
\cline{4-7}
&  &   & $\theta_1$ & $\theta_2$  & $\theta_1$ & $\theta_2$\\
\hline 
.1 & \multirow{2}{*}{$1.5$} & AVE       & 1.50002      &    9.9996e-2  &    1.49894     &    9.9998e-2  \\
\cline{3-7}
& & MAD & 5.4276e-5  &    1.4192e-7 &    1.0683e-3  &    2.4361e-6 \\
\cline{2-7}
& \multirow{2}{*}{$1.7$} & AVE  &  1.50004      &    9.9996e-2  &    1.49893      &    9.9998e-2  \\
\cline{3-7}
& & MAD &  3.5417e-5  &    9.2149e-8 &    1.0902e-3  &    2.4783e-6   \\
\cline{2-7}
& \multirow{2}{*}{$1.9$} & AVE  &  1.50005      &    9.9996e-2   &    1.49893      &    9.9998e-2    \\
\cline{3-7}
& & MAD &   2.6116e-5  &    6.7640e-8   &    1.1004e-3  &    2.4978e-6  \\
\cline{1-7}
1.0 & \multirow{2}{*}{$1.5$} & AVE       & 1.50000      &    9.9997e-2  &    1.49915      &    .10000 \\
\cline{3-7}
& & MAD & 5.3122e-4  &    1.4430e-6  &    1.3131e-3  &    2.9331e-6   \\
\cline{2-7}
& \multirow{2}{*}{$1.7$} & AVE  & 1.50000      &    9.9997e-2 &    1.49909      &    9.9999e-2    \\
\cline{3-7}
& & MAD &   3.2529e-4  &    8.8386e-7  &    1.0788e-3  &    2.4303e-6  \\
\cline{2-7}
& \multirow{2}{*}{$1.9$} & AVE  &  1.50000      &    9.9997e-2 &    1.49905      &    9.9998e-2 \\
\cline{3-7}
& & MAD &  2.3134e-4  &    6.3097e-7  &    1.0094e-3  &    2.3002e-6  \\
\cline{1-7}
\end{tabular}
\end{table}

\begin{table}
\caption{Average estimates and mean absolute deviations of the estimators of frequency and frequency rate for Model 1 when the sample size $n=1000$.} \label{table3}
\begin{tabular}{|c|c|c|c|c|c|c|}
\hline
$\sigma$ & $\alpha$ &  & \multicolumn{2}{|c|}{LSE} & \multicolumn{2}{|c|}{ALSE} \\
\cline{4-7}
&  &   & $\theta_1$ & $\theta_2$  & $\theta_1$ & $\theta_2$\\
\hline 
.1 & \multirow{2}{*}{$1.5$} & AVE       & 1.50005       &    9.9996e-2  &    1.50019      &    9.9996e-2 \\
\cline{3-7}
& & MAD & 1.6431e-5  &    2.2745e-8 &    1.7077e-4  &    2.0331e-7 \\
\cline{2-7}
& \multirow{2}{*}{$1.7$} & AVE  & 1.50006      &    9.9996e-2    &    1.50020      &    9.9996e-2  \\
\cline{3-7}
& & MAD & 1.0062e-5  &    1.3919e-8   &    1.8534e-4  &    2.1688e-7  \\
\cline{2-7}
& \multirow{2}{*}{$1.9$} & AVE  &   1.50006      &    9.9996e-2 &    1.50020      &    9.9996e-2    \\
\cline{3-7}
& & MAD &    6.8519e-6  &    9.4219e-9  &    1.8407e-4  &    2.1479e-7 \\
\cline{1-7}
1.0 & \multirow{2}{*}{$1.5$} & AVE       & 1.50002      &    9.9996e-2  &    1.50013      &    9.9996e-2 \\
\cline{3-7}
& & MAD &  9.7189e-05  &    1.6477e-7  &    1.7441e-4  &    2.3960e-7  \\
\cline{2-7}
& \multirow{2}{*}{$1.7$} & AVE  &  1.50002      &    9.9996e-2  &    1.50014      &    9.9996e-2    \\
\cline{3-7}
& & MAD &  6.6935e-5  &    1.0701e-7 &    1.4911e-4  &    1.8752e-7  \\
\cline{2-7}
& \multirow{2}{*}{$1.9$} & AVE  &  1.50003      &    9.9996e-2  &    1.50014      &    9.9996e-2  \\
\cline{3-7}
& & MAD &   5.2772e-5  &    8.0917e-8 &    1.4178e-4  &    1.7226e-7   \\
\cline{1-7}
\end{tabular}
\end{table} 
 

\begin{table}
\caption{Average estimates and mean absolute deviations of the estimators of frequencies and frequency rates for Model 2 when the sample size $n=250$.}  \label{table4}
\footnotesize{\begin{tabular}{|c|c|c|c|c|c|c||c|c|c|c|}
\hline
$\sigma$ & $\alpha$ &  & \multicolumn{4}{|c|}{LSE} & \multicolumn{4}{|c|}{ALSE} \\
\cline{4-11}
&  &  &   $\theta_{11}$ & $\theta_{12}$  & $\theta_{21}$ & $\theta_{22}$  & $\theta_{11}$ & $\theta_{12}$  & $\theta_{21}$ & $\theta_{22}$ \\
\hline 
.1 & \multirow{2}{*}{$1.5$} & AVE  &  1.50004      &    9.9996e-2  &    2.49990     &   .19999  &  1.50022      &    9.9996e-2  &    2.50739      &   .19997 \\
\cline{3-11}
& & MAD & 3.8164e-5  &    2.4028e-7  &    6.9776e-5  &    4.0953e-7 &  2.4268e-4  &    1.2877e-6  &    7.4010e-3  &    3.5307e-5\\
\cline{2-11}
& \multirow{2}{*}{$1.7$} & AVE  & 1.50005     &    9.9996e-2  &    2.49989      &   .19999  & 1.50018      &    9.9996e-2  &    2.50740      &   .19997  \\
\cline{3-11}
& & MAD  & 2.6862e-5  &    1.6240e-7  &    5.2056e-5  &    2.9084e-7  & 1.8204e-4  &    1.0235e-6  &    7.4186e-3  &    3.5370e-5\\
\cline{2-11}
& \multirow{2}{*}{$1.9$} & AVE   &   1.50005      &    9.9996e-2  &    2.49989      &   .19999 & 1.50015      &    9.9996e-2  &    2.50741      &   .19997\\
\cline{3-11}
& & MAD  & 2.0193e-5  &    1.1984e-7  &    4.1530e-5  &    2.2659e-7  & 1.5003e-4  &    8.9151e-7  &    7.4291e-3  &    3.5407e-5\\
\cline{1-11}
1.0 & \multirow{2}{*}{$1.5$} & AVE   &  1.49999      &    9.9997e-2  &    2.49991      &   .19999 &  1.50049      &    9.9997-2  &    2.50701      &   .19997   \\
\cline{3-11}
& & MAD  &  3.5143e-4  &    2.6203e-6  &    5.9594e-4  &    4.1564e-6   & 2.5948e-3  &    1.1264e-5  &    7.3139e-3  &    3.4540e-5\\
\cline{2-11}
& \multirow{2}{*}{$1.7$} & AVE  &  1.50000      &    9.9997e-2  &    2.49992      &   .19999  & 1.50041      &    9.9996e-2  &    2.50713      &   .19997 \\
\cline{3-11}
& & MAD  &  2.2141e-4  &    1.6372e-6  &    3.6566e-4  &    2.5262e-6  &  1.6068e-3  &    7.0855e-6  &    7.1544e-3  &    3.4295e-5 \\
\cline{2-11}
& \multirow{2}{*}{$1.9$} & AVE   & 1.50000      &    9.9997e-2  &    2.49991      &   .19999  & 1.50040e     &    9.9996e-2  &    2.50726e      &   .19997 \\
\cline{3-11}
& & MAD  &  1.6048e-4  &    1.1801e-6  &    2.6970e-4  &    1.8251e-6  &  1.1269e-3  &    5.0824e-6  &    7.2557e-3  &    3.4733e-5\\
\cline{1-11}
\end{tabular}}
\end{table} 

\begin{table}
\caption{Average estimates and mean absolute deviations of the estimators of frequencies and frequency rates for Model 2 when the sample size $n=500$.}  \label{table5}
\footnotesize{\begin{tabular}{|c|c|c|c|c|c|c||c|c|c|c|}
\hline
$\sigma$ & $\alpha$ &  & \multicolumn{4}{|c|}{LSE} & \multicolumn{4}{|c|}{ALSE} \\
\cline{4-11}
&  &  &   $\theta_{11}$ & $\theta_{12}$  & $\theta_{21}$ & $\theta_{22}$  & $\theta_{11}$ & $\theta_{12}$  & $\theta_{21}$ & $\theta_{22}$ \\
\hline 
.1 & \multirow{2}{*}{$1.5$} & AVE  &  1.50006      &    9.9996e-2  &    2.49991      &   .19999 & 1.49987      &    9.9996e-2  &    2.49989       &   .19999  \\
\cline{3-11}
& & MAD &  9.1188e-6  &    3.5758e-8  &    1.1666e-5  &    4.7214e-8 & 1.22041e-4  &    3.1807e-7  &    7.8160-5  &    2.6177e-7 \\
\cline{2-11}
& \multirow{2}{*}{$1.7$} & AVE  &  1.50006      &    9.9996e-2  &    2.49991      &   .19999  &  1.49988      &    9.9996e-2  &    2.49989      &   .19999  \\
\cline{3-11}
& & MAD  & 7.1482e-6  &    2.4736e-8  &    1.3089e-5  &    3.9897e-8 & 1.0069e-4  &    2.8699e-7  &    6.0609e-5  &    2.3821e-7   \\
\cline{2-11}
& \multirow{2}{*}{$1.9$} & AVE   &  1.50006      &    9.9996e-2  &    2.49991      &   .19999 &  1.49990      &    9.9996e-2  &    2.49989e      &   .19999 \\
\cline{3-11}
& & MAD  &  5.8370e-6  &    1.8806e-8  &    1.3971e-5  &    3.6881e-8 & 8.9066e-5  &    2.6939e-7  &    4.9421e-5  &    2.3137e-7 \\
\cline{1-11}
1.0 & \multirow{2}{*}{$1.5$} & AVE   & 1.50004      &    9.9996e-2  &    2.49990      &   .19999   & 1.49991      &    9.9997e-2  &    2.49993      &   .19999  \\
\cline{3-11}
& & MAD  & 2.9193e-5  &    2.8606e-7  &    3.9704e-5  &    3.8627e-7  &  6.0405e-4  &    1.3624e-6  &    6.8973e-4  &    1.5833e-6 \\
\cline{2-11}
& \multirow{2}{*}{$1.7$} & AVE  &   1.50004      &    9.9996e-2  &    2.49990      &   .19999   &  1.49990      &    9.9997e-2  &    2.49994      &   .19999  \\
\cline{3-11}
& & MAD  &  2.4137e-5  &    1.8561e-7  &    3.0185e-5  &    2.4366e-7  & 3.8765e-4  &    8.7494e-7  &    4.0447e-4  &    9.4990e-7  \\
\cline{2-11}
& \multirow{2}{*}{$1.9$} & AVE   &  1.50004      &    9.9996e-2  &    2.49990      &   .19999   &  1.49989      &    9.9997e-2  &    2.49993      &   .19999 \\
\cline{3-11}
& & MAD  &  2.0650e-5  &    1.3543e-7  &    2.4665e-5  &    1.7835e-7 & 2.9344e-4  &    6.5450e-7  &    2.6731e-4  &    6.5173e-7  \\
\cline{1-11}
\end{tabular}}
\end{table} 

\begin{table}
\caption{Average estimates and mean absolute deviations of the estimators of frequencies and frequency rates for Model 2 when the sample size $n=1000$.}  \label{table6}
\footnotesize{\begin{tabular}{|c|c|c|c|c|c|c||c|c|c|c|}
\hline
$\sigma$ & $\alpha$ &  & \multicolumn{4}{|c|}{LSE} & \multicolumn{4}{|c|}{ALSE} \\
\cline{4-11}
&  &  &   $\theta_{11}$ & $\theta_{12}$  & $\theta_{21}$ & $\theta_{22}$  & $\theta_{11}$ & $\theta_{12}$  & $\theta_{21}$ & $\theta_{22}$ \\
\hline 
.1 & \multirow{2}{*}{$1.5$} & AVE  & 1.50006      &    9.9996e-2  &    2.49991      &   .19999  & 1.50010      &    9.9996e-2  &    2.49983  &   .19999\\
\cline{3-11}
& & MAD & 2.1987e-6  &    5.4090e-9  &    3.4296e-6  &    7.0811e-9  & 5.8498e-5  &    1.5257e-7  &    1.5470e-4  &    2.7713e-7  \\
\cline{2-11}
& \multirow{2}{*}{$1.7$} & AVE  & 1.50006      &    9.9996e-2  &    2.49991      &   .19999 & 1.50008      &    9.9996e-2  &    2.49984      &   .19999   \\
\cline{3-11}
& & MAD  & 1.1567e-6  &    2.5750e-9  &    1.7710e-6  &    2.9147e-9  &  4.4841e-5  &    1.3943e-7  &    1.6391e-4  &    2.8598e-7 \\
\cline{2-11}
& \multirow{2}{*}{$1.9$} & AVE   &  1.50006     &    9.9996e-2  &    2.49991      &   .19999   &  1.50008      &    9.9996e-2  &    2.49983      &   .19999 \\
\cline{3-11}
& & MAD  &  5.9109e-7  &    1.1057e-9  &    7.3657e-7  &    7.9572e-10  &  4.0233e-5  &    1.3410e-7  &    1.7480e-4  &    2.9803e-7 \\
\cline{1-11}
1.0 & \multirow{2}{*}{$1.5$} & AVE   &  1.50006      &    9.9996e-2  &    2.49991      &   .19999 & 1.50012      &    9.9996e-2  &    2.49981      &   .19999  \\
\cline{3-11}
& & MAD  & 5.2064e-6  &    5.1043e-8  &    8.5350e-6  &    6.6623e-8  &  1.2720e-4  &    2.0740e-7  &    1.7314e-4  &    3.1230e-7\\
\cline{2-11}
& \multirow{2}{*}{$1.7$} & AVE  &  1.50006      &    9.9996e-2  &    2.49991      &   .19999  &  1.50013      &    9.9996e-2  &    2.49979      &   .19999\\
\cline{3-11}
& & MAD  &  4.9541e-6  &    3.2181e-8  &    8.0223e-6  &    4.3429e-8  &  1.1081e-4  &    1.8958e-7  &    1.7344e-4  &    3.0105e-7\\
\cline{2-11}
& \multirow{2}{*}{$1.9$} & AVE   &   1.50006      &    9.9996e-2  &    2.49991      &   .19999 & 1.50012      &    9.9996e-2  &    2.49979      &   .19999  \\
\cline{3-11}
& & MAD  &  4.5826e-6  &    2.3501e-8  &    7.6028e-6  &    3.2252e-8  &  9.9376e-5  &    1.8142e-7  &    1.7533e-4  &    3.0316e-7\\
\cline{1-11}
\end{tabular}}
\end{table} 

\section{Concluding Remarks}  \label{conclusions}

In this paper, we consider the multiple chirp signal model observed with additive i.i.d. errors which are heavy tailed.  We propose LS and approximate LS methods to estimate the unknown parameters and prove the strong consistency of the proposed estimators.  No distributional assumption is required to prove the consistency.  We obtain the asymptotic distribution as multivariate symmetric stable distribution when errors are from a symmetric stable distribution.  Although, we have addressed the problem under the assumption of i.i.d. errors, the results can be extended to finite order moving average type.  Another point, we have not considered in the paper, is the estimation of $p$, the number of chirp component.  Some information theoretic criterion or cross validation may be required.  Further work is needed in that direction.

 \section*{Appendix A}
 In this Appendix, we prove Theorem \ref{thm1} and following lemmas are required. 
\begin{lemma} \label{lemma0}
(Lahiri et al. \cite{LKM:2015})  If $(\theta_1, \theta_2) \in (0,\pi)\times (0,\pi)$, then except for a countable number of points, the following
results are true.
\beanno
&& \lim_{n\rightarrow \infty} \frac{1}{n} \sum_{t=1}^n \cos(\theta_1 t +\theta_2 t^2) = \lim_{n\rightarrow \infty} \frac{1}{n} \sum_{t=1}^n \sin(\theta_1 t +\theta_2 t^2) = 0, \\
&& \lim_{n\rightarrow \infty} \frac{1}{n^{k+1}} \sum_{t=1}^n t^k\cos^2(\theta_1 t +\theta_2 t^2) = \lim_{n\rightarrow \infty} \frac{1}{n^{k+1}} \sum_{t=1}^n t^k \sin^2(\theta_1 t +\theta_2 t^2) = \frac{1}{2(k+1)}, \\
&& \lim_{n\rightarrow \infty} \frac{1}{n^{k+1}} \sum_{t=1}^n t^k\cos(\theta_1 t +\theta_2 t^2) \sin(\theta_1 t +\theta_2 t^2) = 0, \;\;\; k=0,1,2. \\
\eeanno
\end{lemma} 
 
 \begin{lemma}  \label{lemma1}
If a sequence of random variables $\{\epsilon(t)\}$ satisfies Assumption \ref{assum1},  then as $n\rightarrow \infty$ and $k \ge 0$
\be
\sup_{\alpha, \beta} \left| \frac{1}{n^{k+1}} \sum_{t=1}^n t^k \epsilon(t) \cos(\alpha t) \cos(\beta t^2) \right| \rightarrow 0, \;\;\; \text{a.s.}  \label{lemma1-eq-1}
\ee
The result is true for all combinations of cosine and sine functions.
\end{lemma}

\noindent {\bf Proof of Lemma \ref{lemma1}:} We prove  \eqref{lemma1-eq-1} for $k=0$ and it follows similarly for $k=1,2, \ldots$.  Define $z(t) = \epsilon(t) I[|\epsilon(t)| \le t^{\frac{1}{1+\delta}}]$.  Then 
\beanno
\sum_{t=1}^\infty P \bls z(t) \ne \epsilon(t)\brs &=& \sum_{t=1}^\infty P \bls |\epsilon(t)| > t^{\frac{1}{1+\delta}} \brs = \sum_{t=1}^\infty  \sum_{2^{t-1} \le n < 2^t}  P \bls |\epsilon(1)| > n^{\frac{1}{1+\delta}} \brs \\
& \le & \sum_{t=1}^\infty  \sum_{2^{t-1} \le n < 2^t} P \bls |\epsilon(1)| > 2^{(t-1)/(1+\delta)} \brs \\
& \le &  \sum_{t=1}^\infty  2^t P \bls |\epsilon(1)| > 2^{(t-1)/(1+\delta)} \brs \\
& \le &  \sum_{t=1}^\infty  2^t \sum_{j=t}^\infty  P \bls 2^{(j-1)/(1+\delta)} \le \epsilon(1) < 2^{(t-1)/(1+\delta)} \brs \\
& \le &  \sum_{j=1}^\infty  P  \bls 2^{(j-1)/(1+\delta)} \le \epsilon(1) < 2^{(t-1)/(1+\delta)} \brs \sum_{t=1}^j 2^t \\
& \le &  C \sum_{j=1}^\infty E|\epsilon(1)|^{1+\delta}  I \bls 2^{(j-1)/(1+\delta)} \le \epsilon(1) < 2^{(t-1)/(1+\delta)} \brs \le E |\epsilon(1)|^{1+\delta} < \infty.
\eeanno
Therefore, $P(z(t) \ne \epsilon(t) \; \text{i.o.}) =0$ and $\{z(t)\}$ and $\{\epsilon(t)\}$ are equivalent sequences.  Hence,
\beanno
&&\sup_{\alpha, \beta} \left| \frac{1}{n} \sum_{t=1}^n  \epsilon(t) \cos(\alpha t) \cos(\beta t^2) \right| \rightarrow 0, \;\;\; \text{a.s.} \\
&\Leftrightarrow &
\sup_{\alpha, \beta} \left| \frac{1}{n} \sum_{t=1}^n  z(t) \cos(\alpha t) \cos(\beta t^2) \right| \rightarrow 0, \;\;\; \text{a.s.}
\eeanno
Write $u(t)=z(t)-E(z(t))$, then for large $n$ 
\be
\sup_{\alpha, \beta} \left| \frac{1}{n} \sum_{t=1}^n  E(z(t)) \cos(\alpha t) \cos(\beta t^2) \right| \le \frac{1}{n} \sum_{t=1}^n  |E(z(t)) | \rightarrow 0.
\ee
Thus, we only need to show that 
\be
\sup_{\alpha, \beta} \left| \frac{1}{n} \sum_{t=1}^n  u(t) \cos(\alpha t) \cos(\beta t^2) \right| \rightarrow 0, \;\;\; \text{a.s.} \label{lemma1-eq2}
\ee
Now, for any fixed $\alpha, \beta \in (0,\pi)$ and $\epsilon >0$, let $0 \le h \le \frac{1}{2 n^{1/(1+\delta)}}$.  Since $|hu(t) \cos(\alpha t) \cos(\beta t^2)| \le \frac{1}{2}$, $e^x \le 1+ x+ 2|x|^{1+\delta}$ for $x \le \frac{1}{2}$ and $E[|u(t)|^{1+\delta} < C$ for some $C >0$,
\beanno
P\Bls  \left| \frac{1}{n} \sum_{t=1}^n  u(t) \cos(\alpha t) \cos(\beta t^2) \right|   \ge  \epsilon \Brs & \le & 2 e^{-hn \epsilon} \prod_{t=1}^n E e^{hu(t) \cos(\alpha t) \cos(\beta t^2) } \\
& \le & 2 e^{-hn \epsilon} \prod_{t=1}^n (1 + C h^{1+\delta}) \\
& \le & 2 e^{-hn \epsilon + 2nCh^{1+\delta}}.
\eeanno
Choose $h=\frac{1}{2 n^{1/(1+\delta)}}$, then for large $n$,
\be
P\Bls  \left| \frac{1}{n} \sum_{t=1}^n  u(t) \cos(\alpha t) \cos(\beta t^2) \right|   \ge  \epsilon \Brs \le 2 e^{-\frac{n\epsilon}{2 n^{\frac{1}{1+\delta}}} + C} = 2 e^{-\frac{\epsilon}{2} n^{\frac{\delta}{1+\delta}}}.
\ee
Take $K = n^6$.  Choose $K$ points $(\alpha_1, \beta_1), \ldots, (\alpha_K, \beta_K)$ such that for any
point $(\alpha, \beta)$ in $(0, \pi) \times (0, \pi)$, we have a point $(\alpha_j, \beta_j)$ satisfying
$$
|\alpha_j - \alpha| \le \frac{\pi}{n^3} \hspace{.25in} \hbox{and} \hspace{.15in}
 |\beta_j - \beta| \le \frac{\pi}{n^3}.
$$
Now Taylor series expansion can be used to bound $|\cos(\beta t^2)
-\cos(\beta_j t^2)| \le t^2 |\beta-\beta_j|$ and $|\cos(\alpha t) -\cos(\alpha_j t)| \le
|t||\alpha-\alpha_j|$, therefore,
\beanno 
& & \left | \frac{1}{n} \sum_{t=1}^n u(t) \left \{
\cos(\alpha t) \cos(\beta t^2) - \cos(\alpha_j t)
\cos(\beta_j t^2) \right \} \right |  \\
& & \le  \left | \frac{1}{n} \sum_{t=1}^n u(t) \cos(\alpha t) \left \{ \cos(\beta t^2) -
\cos(\beta_j t^2)
 \right \} \right | +
 \left | \frac{1}{n} \sum_{t=1}^n u(t) \left \{ \cos(\beta_j t^2)(\cos(\alpha t) - \cos(\alpha_j t)
\right \} \right |   \\
& & \hspace{3.95in} \mbox{(using Taylor series expansion)}     \\
& & \le C \left [ \frac{1}{n} \sum_{t=1}^n t^{\frac{1}{1+\delta}} t^2 \frac{\pi}{n^3} +
\frac{1}{n} \sum_{t=1}^n t^{\frac{1}{1+\delta}} t \frac{\pi}{n^3} \right ]  \le
C \left [ \frac{\pi}{n^{\frac{\delta}{1+\delta}}} +  \frac{\pi}{n^{\frac{1+2\delta}{1+\delta}}} \right ]
{\longrightarrow} \hspace{.05in} 0  \hspace{.25in} \hbox{as} \hspace{.15in} n \rightarrow \infty.
\eeanno
Therefore for large $n$, we have
\beanno
P \left [ \sup_{\alpha, \beta} \left | \frac{1}{n} \sum_{t=1}^n u(t) \cos(\alpha t) \cos(\beta t^2)
\right | \ge 2 \epsilon \right ] & \le &
P \left [ \max_{j \le n^6} \left | \frac{1}{n} \sum_{t=1}^n u(t) \cos(\alpha_j t)
\cos(\beta_j t^2)
\right | \ge \epsilon \right ]  \\
& \le & 2 n^6 e^{-\frac{\epsilon}{2} n^{\frac{\delta}{1+\delta}}}.  \\
\eeanno
Since $\sum_{n=1}^{\infty} 2 n^6 e^{-\frac{\epsilon}{2} n^{\frac{\delta}{1+\delta}}} < \infty$,
therefore because of Borel-Cantelli lemma \eqref{lemma1-eq2} holds true and that proves the lemma.  \qed

\begin{lemma}  \label{lemma2}
Write 
\beanno
S_{c} &=& \{\ve{\xi} \in {\ve \Omega}: ~\ve{\xi} = 
(A, B, \theta_1, \theta_2), |{\ve\xi} - {\ve\xi}^0| \ge 3c\}
\eeanno
where ${\ve \Omega} = (-M, M)\times (-M, M) \times (0,\pi)\times (0,\pi)$.  
If for any $c > 0$,
\be
\underline{\lim} \inf_{\ve{\theta} \in S_{c}} \frac{1}{n} \left 
[ Q(\ve{\xi}) - Q(\ve{\xi}^0) \right ] > 0 \hspace{.25in} a.s.,
         \label{lemma2-eq-1}
\ee
then $\wh{\ve{\xi}}$, which minimizes $Q({\ve\xi})$, is a strongly 
consistent estimator of $\ve{\xi}^0$.  
\end{lemma}

\noindent {\bf Proof of Lemma \ref{lemma2}:} The proof can be obtained along the same line as the proof of Lemma 1 of \cite{Wu:1981}.

\noindent {\bf Proof of Theorem \ref{thm1}:}  Write 
$
\frac{1}{n} \left [ Q({\ve \xi}) - Q({\ve \xi}^0) \right ] = f_1({\ve
  \xi}) + f_2({\ve \xi}), 
$
where
\beanno
f_1({\ve \xi})  & = & \frac{1}{t} \sum_{t=1}^n \left [ A^0 \cos(\theta_1^0 t +
  \theta_2^0 t^2) -  
A \cos(\theta_1 t + \theta_2 t^2) \right .  \\
& & \ \ \ \ \ \ \ \ \ 
\left .  + B^0 \sin(\theta_1^0 t + \theta_2^0 t^2) - 
B \sin(\theta_1 t + \theta_2 t^2) \right ]^2,
\eeanno
\beanno
f_2({\ve \xi}) & = & \frac{2}{n} \sum_{t=1}^n X(t)\left [ A^0 \cos(\theta_1^0
  t + \theta_2^0 t^2) -  
A \cos(\theta_1 t + \theta_2 t^2) \right . \\
& & \ \ \ \ \ \ \ \ \ 
\left . + B^0 \sin(\theta_1^0 t + \theta_2^0 t^2) - 
B \sin(\theta_1 t + \theta_2 t^2) \right ].
\eeanno
Using Lemma \ref{lemma1}, it follows that $\sup_{{\ve\xi} \in S_{c}}  |f_2({\ve\xi})| \rightarrow 0$, a.s.  It remains to prove that \\
$\liminf \inf_{{\ve\xi}\in S_{c}} f_1({\ve\xi}) > 0$ a.s.

Consider the following sets:
\beanno
S_{c,1} & = & \left \{{\ve \xi}: {\ve \xi} = (A,B,\theta_1,\theta_2),
|A-A^0| \ge c
\right \}, \\
S_{c,2} & = & \left \{{\ve \xi}: {\ve \xi} =(A,B,\theta_1,\theta_2),
|B-B^0| \ge c  
\right \}, \\
S_{c,3} & = & \left \{{\ve \xi}: {\ve \xi} = (A,B,\theta_1,\theta_2),
|{\ve \theta}-{\ve \theta}^0| \ge c \right \}, {\ve\theta} = (\theta_1, \theta_2)
\eeanno
Note that 
$S_{c} \subset S_{c,1} \cup S_{c,2} \cup S_{c,3} = S
\hspace{.2in} \hbox{(say)}$.  Therefore,
\be
\underline{\lim} \inf_{{\ve \xi} \in S_{c}} \frac{1}{n} 
\left [ Q({\ve \xi}) - Q({\ve \xi^0}) \right ] \ge
\underline{\lim} \inf_{{\ve \xi} \in S} \frac{1}{n} 
\left [ Q({\ve \xi}) - Q({\ve \xi^0}) \right ].      \label{inequality}
\ee  
Hence, 
\beanno
& & \underline{\lim} \inf_{{\ve \xi} \in S_{c,1}} \frac{1}{n} 
\left [ Q({\ve \xi}) - Q({\ve \xi^0}) \right ] = \underline{\lim}
\inf_{{\ve \xi} \in S_{c,1}} f_1({\ve  
\xi})  \\
& & = \underline{\lim} \inf_{|A - A^0| \ge c} \frac{1}{n} \sum_{t=1}^n \left [
  A^0 \cos(\theta_1^0 t + \theta_2^0 t^2)  
- A \cos(\theta_1 t + \theta_2 t^2) + \right .   \\ 
     &  &  \hspace{1.5in} \left .  B^0 \sin(\theta_1^0 t + \theta_2^0 t^2) - 
B \sin(\theta_1 t + \theta_2 t^2) \right ]^2    \\
& & = \lim_{n \rightarrow \infty} \inf_{|A - A^0| \ge c} \frac{1}{n} \sum_{t=1}^n \cos^2 (\theta_1^0 t + \theta_2^0 t^2) (A - A^0)^2  \\
& & \ge c^2 \lim_{n \rightarrow \infty} \frac{1}{n} \sum_{t=1}^n \cos^2 (\theta_1^0 t + \theta_2^0 t^2) > 0.
\eeanno
One can proceed along the same lines  for $S_{c,2}$ and $S_{c,3}$ and that proves the theorem.  \qed

\section*{Appendix B}

In this Appendix, we provide the proofs of consistency results related to ALSEs.
\begin{lemma}  \label{lemma3}
Let $\wt{\ve\theta}=(\wt{\theta}_1, \wt{\theta}_2)$ be an estimate of ${\ve\theta}^0 = (\theta_1^0, \theta_2^0)$ that maximizes $I(\theta_1, \theta_2)$ and for any $\epsilon >0$, let $S_\epsilon =\blb {\ve\theta}: |\wt{\ve\theta} - {\ve\theta}^0| > \epsilon \brb$ for some fixed ${\ve\theta}^0 \in (0,\pi)\times (0,\pi)$.  If for any $\epsilon >0$,
\be
\overline{\lim}_{n\rightarrow \infty} \sup_{S_\epsilon} \frac{1}{n} \bls I(\theta_1, \theta_2) - I(\theta_1^0, \theta_2^0) \brs < 0, \;\; \text{a.s.}  \label{lemma3_eq}
\ee
then as $n\rightarrow \infty$, $\wt{\ve\theta} \rightarrow {\ve\theta}^0$ a.s., that is, $\wt{\theta}_1 \rightarrow \theta_1^0$ and $\wt{\theta}_2 \rightarrow \theta_2^0$ a.s.
\end{lemma}

\noindent {\bf Proof of Lemma \ref{lemma3}:} We write $\widetilde{\ve\theta}$ by $\widetilde{\ve\theta}_n$ and $I(\theta_1, \theta_2)=I({\ve\theta})$ by $I_n({\ve\theta})$ to emphasize that these quantities depend on $n$.  Suppose \eqref{lemma3_eq} is true but $\widetilde{\ve\theta}_n$ does not converges to ${\ve\theta}^0$ as $n\rightarrow \infty$. Then, there exists an $\epsilon >0$ and a subsequence $\{n_k\}$ of $\{n\}$ such that $|\wt{\ve\theta}_{n_k} - {\ve\theta}^0| > \epsilon$ for $k=1,2,\ldots$.  Therefore, $\wt{\ve\theta}_{n_k} \in S_\epsilon$ for all $k=1,2,\ldots$.  By definition, $\wt{\ve\theta}_{n_k}$ is the ALSE of ${\ve\theta}^0$ and hence maximizes $I_{n_k}({\ve\theta})$ when $n=n_k$.  This implies that 
\beanno
I_{n_k}(\wt{\ve\theta}_{n_k}) \ge I_{n_k}({\ve\theta}^0) \Rightarrow \frac{1}{n_k} \Bls I_{n_k}(\wt{\ve\theta}_{n_k}) - I_{n_k}({\ve\theta}^0) \Brs  \ge 0.
\eeanno
Therefore, $\ds \overline{\lim}_{n\rightarrow \infty} \sup_{{\ve\theta}_{n_k} \in S_\epsilon} \frac{1}{n_k} \bls I_{n_k}(\wt{\ve\theta}_{n_k}) - I_{n_k}({\ve\theta}^0) \brs \ge 0$, which contradicts the inequality \eqref{lemma3_eq}.  Hence, the result follows. \qed

\begin{lemma}  \label{lemma4}
If the error process $\{e(t)\}$ satisfies Assumption \ref{assum1} and the amplitudes $A^0$ and $B^0$ satisfy Assumption \ref{assum6}, then the estimator $\wt{\ve\theta}$ of ${\ve\theta}^0$ which maximizes $I({\ve\theta})$, as defined in \eqref{i-eq}, is a strongly consistent estimator of ${\ve\theta}^0$.
\end{lemma}

\noindent {\bf Proof of Lemma \ref{lemma4}:} In this proof, we write $\theta_1 t + \theta_2 t^2=h({\ve\theta};t)$.  Consider
\beanno
\frac{1}{n} \bls I({\ve\theta}) - I({\ve\theta}^0) \brs &=& \frac{1}{n} \Bls \frac{2}{n}\left| \sum_{t=1}^n y(t) e^{- i(\theta_1 t + \theta_2 t^2)} \right|^2  - 
\frac{2}{n}\left| \sum_{t=1}^n y(t) e^{- i(\theta_1^0 t + \theta_2^0 t^2)} \right|^2 \Brs \\
&=& \frac{2}{n^2}\left[ \Blp \sum_{t=1}^n y(t) \cos(h({\ve\theta};t)) \Brp^2 + \Blp\sum_{t=1}^n y(t) \sin(h({\ve\theta};t)) \Brp^2 \right. \\
& &  \left. - \Blp \sum_{t=1}^n y(t) \cos(h({\ve\theta}^0;t)) \Brp^2 - \Blp \sum_{t=1}^n y(t) \sin(h({\ve\theta}^0;t)) \Brp^2 \right] \\
&=& \frac{2}{n^2}\left[ \Blp \sum_{t=1}^n \Blb A^0 \cos(h({\ve\theta}^0;t)) + B^0 \sin(h({\ve\theta}^0;t))
+ e(t) \Brb \cos(h({\ve\theta};t)) \Brp^2 \right. \\
& & + \Blp \sum_{t=1}^n \Blb A^0 \cos(h({\ve\theta}^0;t)) + B^0 \sin(h({\ve\theta}^0;t))
+ e(t) \Brb \sin(h({\ve\theta};t)) \Brp^2  \\
& & - \Blp \sum_{t=1}^n \Blb A^0 \cos(h({\ve\theta}^0;t)) + B^0 \sin(h({\ve\theta}^0;t))
+ e(t) \Brb \cos(h({\ve\theta}^0;t)) \Brp^2 \\
& & \left. - \Blp \sum_{t=1}^n \Blb A^0 \cos(h({\ve\theta}^0;t)) + B^0 \sin(h({\ve\theta}^0;t))
+ e(t) \Brb \sin(h({\ve\theta}^0;t)) \Brp^2  \right].
\eeanno
Using identities given in Lemma \ref{lemma0}, we have 
\beanno
& & \overline{\lim}_{n\rightarrow \infty} \sup_{S_\epsilon} \frac{1}{n} \bls I({\ve\theta}) - I({\ve\theta}^0) \brs \\
&=&  2 \overline{\lim}_{n\rightarrow \infty} \sup_{S_\epsilon} \left[ \Blp \frac{A^0}{n} \sum_{t=1}^n \cos(h({\ve\theta}^0;t)) \cos(h({\ve\theta};t)) \Brp^2
+ \Blp \frac{B^0}{n} \sum_{t=1}^n \sin(h({\ve\theta}^0;t)) \sin(h({\ve\theta};t)) \Brp^2  \right. \\
& & \left. \hspace{1in} - \Blp \frac{A^0}{n} \sum_{t=1}^n \cos^2 (h({\ve\theta}^0;t)) \Brp^2 - \Blp \frac{A^0}{n} \sum_{t=1}^n \sin^2 (h({\ve\theta}^0;t)) \Brp^2  \right] \\
&=& -2 {A^0}^2 . \frac{1}{4} - 2 {B^0}^2 . \frac{1}{4} = -\frac{1}{2} ({A^0}^2 + {B^0}^2), \;\; \text{a.s.}
\eeanno
Therefore, using Lemma \ref{lemma3}, the result follows.  \qed

\begin{lemma}  \label{lemma5}
Under the same assumptions as in Lemma \ref{lemma4}, $n(\wh{\theta}_1 - \theta_1^0) \rightarrow 0$ and $n^2(\wh{\theta}_2 - \theta_2^0) \rightarrow 0$, a.s. as $n\rightarrow \infty$. 
\end{lemma}

\noindent {\bf Proof of Lemma \ref{lemma5}:} Let $I'({\ve\theta}) = (\frac{\partial I({\ve\theta}) }{\partial \theta_1}, \frac{\partial I({\ve\theta}) }{\partial \theta_2})$ and $I''({\ve\theta})$ be the $2\times 2$ matrix of second derivatives of $I({\ve\theta})$ with respect to ${\ve\theta}$.  Applying  similar steps applied to $Q({\ve\xi})$ in section \ref{asym-dist-single-chirp}, expand $I'({\ve\theta})$ at $\wt{\ve\theta}$ around ${\ve\theta}^0$ using Taylor series expansion
\be
I'(\wt{\ve\theta}) - I'({\ve\theta}^0) = (\wt{\ve\theta} - {\ve\theta}^0) I''(\bar{\ve\theta}),  \label{lemma5-eq1}
\ee
where $\bar{\ve\theta}$ is a point on the line joining $\wt{\ve\theta}$ and ${\ve\theta}^0$.  Define a diagonal matrix of order $2$ as ${\bf D}=\text{diag}\blb \frac{1}{n}, \frac{1}{n^2} \brb$.,  then \eqref{lemma5-eq1} can be written as 
\be
(\wt{\ve\theta} - {\ve\theta}^0) {\bf D}^{-1} = \Bls \frac{1}{n} I'({\ve\theta}^0) {\bf D} \Brs \Bls \frac{1}{n}{\bf D} I''(\bar{\ve\theta}){\bf D} \Brs^{-1}
\label{lemma5-eq2}
\ee
because $I'(\wt{\ve\theta}) =0$ ($\wt{\ve\theta}$ maximizes $I({\ve\theta})$). Now note that 
\be
\lim_{n\rightarrow \infty}  \Bls \frac{1}{n}{\bf D} I''(\bar{\ve\theta}){\bf D} \Brs = \lim_{n\rightarrow \infty}  \Bls \frac{1}{n}{\bf D} I''({\ve\theta}^0){\bf D} \Brs = -\frac{1}{{A^0}^2 + {B^0}^2} \left(\begin{matrix}
\frac{1}{12}  & \frac{1}{12}  \\
\frac{1}{12}  & \frac{4}{45}   \label{lemma5-eq3}
\end{matrix} \right)
\ee
using Lemma \ref{lemma0} and additionally using Lemma \ref{lemma1}, we have 
\be
\Bls \frac{1}{n} I'({\ve\theta}^0) {\bf D} \Brs  \rightarrow {\bf 0}, \;\; \text{a.s.} \;\; \text{as} \;\; n \rightarrow \infty.  \label{lemma5-eq4}
\ee
Using \eqref{lemma5-eq3} and \eqref{lemma5-eq4} in \eqref{lemma5-eq2}, we have $(\wt{\ve\theta} - {\ve\theta}^0) {\bf D}^{-1} \rightarrow 0$ a.s. as $n \rightarrow \infty$ which implies that $n(\wh{\theta}_1 - \theta_1^0) \rightarrow 0$ and $n^2(\wh{\theta}_2 - \theta_2^0) \rightarrow 0$, a.s. as $n\rightarrow \infty$.\qed

\begin{lemma} \label{lemma6}
Under the same assumptions as in Lemma \ref{lemma4}, $\wt{A}$ and $\wt{B}$ are strongly consistent estimator of $A^0$ and $B^0$.
\end{lemma}

\noindent {\bf Proof of Lemma \ref{lemma6}:}  Expanding $\cos(\wt{\theta}_1 t + \wt{\theta}_2 t^2)=\cos(h(\wt{\ve\theta};t)$  around ${\ve\theta}^0$ upto the first order term using multivariate Taylor series
\beanno
\wt{A} &=& \frac{2}{n} \sum_{t=1}^n y(t) \cos(\widetilde{\theta}_1 t + \widetilde{\theta}_2 t^2) = \frac{2}{n} \sum_{t=1}^n y(t) \cos(h(\wt{\ve\theta};t)) \\
&=& \frac{2}{n} \sum_{t=1}^n \Bls A^0 \cos(h({\ve\theta}^0;t)) + B^0 \sin(h({\ve\theta}^0;t))
+ e(t) \Brs \Bls \cos(h({\ve\theta}^0;t)) \\ 
& & \hspace{1in}- t(\wt{\theta}_1 - \theta_1^0) \sin(h(\bar{\ve\theta};t)) - t^2(\wt{\theta}_2 - \theta_2^0) \sin(h(\bar{\ve\theta};t)) \Brs \\
&=& \frac{2A^0}{n} \sum_{t=1}^n \cos^2(h({\ve\theta}^0;t)) + \frac{2B^0}{n} \sum_{t=1}^n \sin(h({\ve\theta}^0;t)) \cos(h({\ve\theta}^0;t)) +
\frac{2}{n} \sum_{t=1}^n e(t) \cos(h({\ve\theta}^0;t)) \\
&& - 2 n(\wt{\theta}_1 - \theta_1^0) \frac{1}{n^2} \sum_{t=1}^n t \Bls A^0 \cos(h({\ve\theta}^0;t)) \sin(h(\bar{\ve\theta};t)) \\
&& \hspace{2in}+ B^0 \sin(h({\ve\theta}^0;t)) \sin(h(\bar{\ve\theta};t)) + e(t) \sin(h(\bar{\ve\theta};t)) \Brs \\
&& - 2 n^2(\wt{\theta}_2 - \theta_2^0) \frac{1}{n^3} \sum_{t=1}^n t^2 \Bls A^0 \cos(h({\ve\theta}^0;t)) \sin(h(\bar{\ve\theta};t)) \\
&& \hspace{2in} + B^0 \sin(h({\ve\theta}^0;t)) \sin(h(\bar{\ve\theta};t)) + e(t) \sin(h(\bar{\ve\theta};t)) \Brs \\
& \rightarrow & A^0 \;\; \text{a.s.} \;\; \text{as} \;\; n \rightarrow \infty,
\eeanno
using Lemmas \ref{lemma0}, \ref{lemma1} and \ref{lemma5}.  Similarly expanding $\sin(\wt{\theta}_1 t + \wt{\theta}_2 t^2)$, it can be shown that $\wt{B} \rightarrow B^0$ a.s.    \qed

\section*{Appendix C}  

In this Appendix, we show for model \eqref{single-chirp-model} that the asymptotic distribution of ALSE $\wt{\ve\xi}$ of ${\ve\xi}^0$ is same as the asymptotic distribution of the LSE $\wh{\ve\xi}$ under Assumption \ref{assum2}.
\begin{lemma}  \label{lemma7}
If $(\theta_1, \theta_2) \in (0,\pi)\times (0,\pi)$, then except for a countable number of points, the following
results are true.
\beanno
&& \lim_{n\rightarrow \infty} \frac{1}{\sqrt{n}} \sum_{t=1}^n \cos(\theta_1 t +\theta_2 t^2) = \lim_{n\rightarrow \infty} \frac{1}{\sqrt{n}} \sum_{t=1}^n \sin(\theta_1 t +\theta_2 t^2) = 0, \\
&& \lim_{n\rightarrow \infty} \frac{1}{n^{\frac{3}{2}}} \sum_{t=1}^n t \cos(\theta_1 t +\theta_2 t^2) = \lim_{n\rightarrow \infty} \frac{1}{n^{\frac{3}{2}}} \sum_{t=1}^n t\sin(\theta_1 t +\theta_2 t^2) = 0, \\
&& \lim_{n\rightarrow \infty} \frac{1}{n^{\frac{5}{2}}} \sum_{t=1}^n t^2 \cos(\theta_1 t +\theta_2 t^2) = \lim_{n\rightarrow \infty} \frac{1}{n^{\frac{5}{2}}} \sum_{t=1}^n t^2\sin(\theta_1 t +\theta_2 t^2) = 0.
\eeanno
\end{lemma}

\noindent {\bf Proof of Lemma \ref{lemma7}:}  Refer to Grover et al. \cite{GKM:2018}

\noindent {\bf Proof of Theorem \ref{thm33}:} Let $Q({\ve\xi})$ be the residual sum of squares defined in \eqref{sum-square-eq}.  Then 
\beanno
\frac{1}{n} Q({\ve\xi}) &=& \frac{1}{n} \sum_{t=1}^n \Blp y(t) - A \cos(\theta_1 t + \theta_2 t^2) - B \sin(\theta_1 t + \theta_2 t^2) \Brp^2  \\
&=& \frac{1}{n} \sum_{t=1}^n y^2(t) -\frac{2}{n} \sum_{t=1}^n y(t) \blp A \cos(\theta_1 t + \theta_2 t^2) + B \sin(\theta_1 t + \theta_2 t^2) \brp \\
&&  \hspace{1in}+  \frac{1}{n} \sum_{t=1}^n \blp A \cos(\theta_1 t + \theta_2 t^2) + B \sin(\theta_1 t + \theta_2 t^2) \brp^2 \\
&=& \frac{1}{n} \sum_{t=1}^n y^2(t) -\frac{2}{n} \sum_{t=1}^n y(t) \blp A \cos(\theta_1 t + \theta_2 t^2) + B \sin(\theta_1 t + \theta_2 t^2) \brp  + 
\frac{1}{2} (A^2 + B^2) + o(1)\\
&=& C - \frac{1}{n} J({\ve\xi}) + o(1),
\eeanno
where $ C=\frac{1}{n} \sum_{t=1}^n y^2(t) $ and $J({\ve\xi}) = -\frac{2}{n} \sum_{t=1}^n y(t) \blp A \cos(\theta_1 t + \theta_2 t^2) + B \sin(\theta_1 t + \theta_2 t^2) \brp + \frac{1}{2} (A^2 + B^2)$.   Now compute the elements of $\frac{1}{n} J'({\ve\xi})$ at ${\ve\xi}^0$ where $J'({\ve\xi}) = (\frac{\partial J({\ve\xi})}{\partial A}, \frac{\partial J({\ve\xi})}{\partial B}, \frac{\partial J({\ve\xi})}{\partial \theta_1}, \frac{\partial J({\ve\xi})}{\partial \theta_2})$.  Using Lemmas \ref{lemma0}, \ref{lemma1} and \ref{lemma7}, we observe that 
\beanno
\left. \frac{1}{n} \frac{\partial J({\ve\xi})}{\partial A}\right|_{{\ve\xi}^0} &=& \frac{2}{n} \sum_{t=1}^n e(t) \cos(\theta_1^0 t + \theta_2^0 t^2) + o\blp\frac{1}{\sqrt{n}}\brp, \\
\left. \frac{1}{n} \frac{\partial J({\ve\xi})}{\partial B}\right|_{{\ve\xi}^0} &=& \frac{2}{n} \sum_{t=1}^n e(t) \sin(\theta_1^0 t + \theta_2^0 t^2) + o\blp\frac{1}{\sqrt{n}}\brp, \\
\left. \frac{1}{n} \frac{\partial J({\ve\xi})}{\partial \theta_1}\right|_{{\ve\xi}^0} &=& \frac{2}{n} \sum_{t=1}^n t e(t) \bls A^0 \sin(\theta_1^0 t + \theta_2^0 t^2) - B^0 \cos(\theta_1^0 t + \theta_2^0 t^2) \brs  +o\blp \sqrt{n}\brp, \\
\left. \frac{1}{n} \frac{\partial J({\ve\xi})}{\partial \theta_2}\right|_{{\ve\xi}^0} &=& \frac{2}{n} \sum_{t=1}^n t^2 e(t) \bls A^0 \sin(\theta_1^0 t + \theta_2^0 t^2) - B^0 \cos(\theta_1^0 t + \theta_2^0 t^2) \brs  +o\blp n \sqrt{n}\brp.
\eeanno
Comparing with $\frac{1}{n} Q'({\ve\xi}^0)$, we have 
\[
\frac{1}{n} Q'({\ve\xi}^0) {\bf D}_1 = - \frac{1}{n}  J'({\ve\xi}^0) {\bf D}_1 + \Blp o\blp\frac{1}{\sqrt{n}}\brp, \;\; o\blp\frac{1}{\sqrt{n}}\brp, \; \; o\blp \sqrt{n}\brp, \;\; o\blp n \sqrt{n}\brp \Brp {\bf D}_1,
\]
where ${\bf D}_1$ is same as defined in section \ref{asym-dist-single-chirp}.  This implies that 
\be
\lim_{n\rightarrow \infty} Q'({\ve\xi}^0) {\bf D}_1 = - \lim_{n\rightarrow \infty} J'({\ve\xi}^0) {\bf D}_1,  \label{thm33_eq1}
\ee
because $\ds \lim_{n\rightarrow\infty} \blp o\blp\sqrt{n}\brp, \;\; o\blp\sqrt{n}\brp, \; \; o\blp n\sqrt{n}\brp, \;\; o\blp n^2 \sqrt{n}\brp \Brp {\bf D}_1 = {\bf 0}$ for $1+\delta < \alpha <2$, $0<\delta < 1$.  

We note that ALSEs $\wt{A}$ and $\wt{B}$ are expressed as functions of $(\theta_1, \theta_2)$ and replacing $A$ and $B$ with $\wt{A}$ and $\wt{B}$ in $J({\ve\xi})$, it is observed that 
\[
J(\wt{A}, \wt{B}, \theta_1, \theta_2) = I(\theta_1, \theta_2).
\]
Therefore, the estimator of ${\ve\xi}^0$ that maximizes $J({\ve\xi})$, is equivalent to $\wt{\ve\xi}$, the ALSE of ${\ve\xi}^0$.  The estimating equation for ALSE $\wt{\ve\xi}$ in terms of $J({\ve\xi})$ is 
\bea
&& \left ( \wt{\ve{\xi}} - \ve{\xi}^0 \right ) J''(\bar{\ve{\xi}}) 
= -  J'(\ve{\xi}^0)  \nonumber  \\
&\Rightarrow & \left ( \wt{\ve{\xi}} - \ve{\xi}^0 \right ) {\bf D}_2^{-1}
= - \left [ J'(\ve{\xi}^0) {\bf D}_1 \right ] \left [ {\bf D}_2
J''(\bar{\ve{\xi}}) {\bf D}_1 \right ]^{-1}.  \label{thm33_eq11}
\eea
Now proceeding similarly as in case of $Q''(\ve\xi)$ is section \ref{sec3}, we have for $J''(\ve\xi)$
\be
\lim_{n \rightarrow \infty} \left [ {\bf D}_2
J''(\bar{\ve{\xi}}) {\bf D}_1 \right ] =
\lim_{n \rightarrow \infty} \left [ {\bf D}_2
J''(\ve{\xi}^0) {\bf D}_1 \right ] = -{\ve \Gamma} = \lim_{n \rightarrow \infty} \left [ {\bf D}_2
Q''(\ve{\xi}^0) {\bf D}_1 \right ]  \label{thm33_eq2}
\ee
where ${\ve\Gamma}$ is defined in \eqref{gamma_eq}.  Hence, using \eqref{thm33_eq1} and \eqref{thm33_eq2} in \eqref{thm33_eq11}, we have for large $n$, 
\beanno
\left ( \wt{\ve{\xi}} - \ve{\xi}^0 \right ) {\bf D}_2^{-1} &=& - \left [ J'(\ve{\xi}^0) {\bf D}_1 \right ] \left [ {\bf D}_2
J''(\bar{\ve{\xi}}) {\bf D}_1 \right ]^{-1} \\
&=& - \left [ Q'(\ve{\xi}^0) {\bf D}_1 \right ] \left [ {\bf D}_2
Q''(\bar{\ve{\xi}}) {\bf D}_1 \right ]^{-1} \\
&=& \left ( \wh{\ve{\xi}} - \ve{\xi}^0 \right ) {\bf D}_2^{-1}.
\eeanno
Therefore, it follows that the LSE, $\wh{\ve\xi}$ and ALSE, $\wt{\ve\xi}$ of ${\ve\xi}^0$ are asymptotically equivalent in distribution and asymptotic distribution of $\wt{\ve\xi}$ is same as $\wh{\ve\xi}$.

\end{document}